\documentclass{aa}  
\usepackage{graphicx}
\usepackage{txfonts}
\usepackage{amssymb}
\DeclareMathOperator{\Erf}{Erf}
\let\oldphi\phi \let\phi\varphi \let\varphi\oldphi
\begin{document} 
   \title{Relativistic dynamical friction in stellar systems}
   \author{Caterina Chiari
          \inst{1}
          \and
          Pierfrancesco Di Cintio\inst{2,3,4}\fnmsep
          }
   \institute{Dipartimento di Scienze Fisiche, Informatiche e Matematiche, Universit\`a di Modena e Reggio Emilia, Via Campi 213/A, I-41125 Modena, Italy\\
              \email{caterina.chiari@unimore.it}
         \and
             CNR - ISC, Via Madonna del piano 10, I-50019 Sesto Fiorentino, Italy\\
         \and
            INAF - Osservatorio Astrofisico di Arcetri, Largo Enrico Fermi 5, I-50125 Firenze, Italy\\
        \and
           INFN -  Sezione di Firenze, via G.\ Sansone 1, I-50019 Sesto Fiorentino, Italy\\
             \email{pierfrancesco.dicintio@cnr.it}
             }
   \date{Received September xx, 2022; accepted March xx, 2022}
%%%%%%%%%%%%%%%%%%%%%%%%%%%%%%%%%%%%%%%%%%%%%%%%%%%%%%
  \abstract
   {}
   {We extend the classical formulation of the dynamical friction effect on a test star by Chandrasekhar to the case of relativistic velocities and velocity distributions also accounting for post-Newtonian corrections to the gravitational force.}
   {The original kinetic framework is revised and used to construct a special-relativistic dynamical friction formula where the relative velocities changes in subsequent encounters are added up with Lorentz transformation and the velocity distribution of the field stars accounts for relativistic velocities. Furthermore, a simple expression is obtained for systems where the post-Newtonian correction on the gravitational forces become relevant even at non-relativistic particle velocities. Finally, using a linearized Lagrangian we derive another expression for the dynamical friction expression in a more compact form than that of Lee (1969).}
   {Comparing our formulation with the classical one, we observe that a given test particle suffers a slightly stronger drag when moving through a distribution of field stars with relativistic velocity distribution. Vice versa, a purely classical treatment of a system where post-Newtonian (PN) corrections should be included, over estimates the effect of dynamical friction at low test particle velocity, regardless of the form of velocity distribution. Finally, a first order PN dynamical friction covariant formulation is less strong than its classical counterpart at small velocities but much higher for large velocities over a broad range of mass ratios.}
   {}
%%%%%%%%%%%%%%%%%%%%%%%%%%%%%%%%5
   \maketitle
%-------------------------------------------------------------------
\section{Introduction}
Dynamical friction (hereafter DF) is an important physical phenomenon, with several consequences in stellar dynamics (and in plasma physics). It can be qualitatively thought of as the slowing-down of a test particle of mass $M$, moving at $\mathbf{v}_{\rm T}$ in a background of field particles of mass $m$, mean number density $n$ and velocity distribution $f(\mathbf{v}_{\rm F})$, due to the cumulative effect of their long-range gravitational (or Coulomb) interactions.\\
\indent An analytical estimation of the DF was evaluated for the first time for stellar systems by \cite{Chandra:43}, who found that $M$ must experience a slowing down along its initial direction of propagation as
\begin{equation}\label{dfchandra}
\frac{d\textbf{v}_{\rm T}}{dt}= -4\pi G^2nm(M+m)\log\Lambda\frac{ \Xi(v_{\rm T})  }{v^3_{\rm T}} \textbf{v}_{\rm T}.
\end{equation}
In the equation above, $G$ is the gravitational constant, $\log\Lambda$ is the so-called Coulomb logarithm (in analogy with the analogous quantity in plasma physics, \citealt{1965pfig.book.....S}) of the ratio $\Lambda$ of the maximum and minimum impact parameters $b_{\rm max}$ and $b_{\rm min}$ and
\begin{equation}\label{velvol}
\Xi(v_{\rm T})\equiv 4\pi\int_{0}^{v_{\rm T}} f(v_{\rm F})v_{\rm F}^2 dv_{\rm F}
\end{equation}
is the fractional velocity volume function.\\
\indent The process of DF is crucial for the evolution of collisional systems (see e.g. \citealt{BinTre:87}) from the large scales of galaxies clusters (\citealt{1975ApJ...202L.113O,1976ApJ...204..642R,1976ApJ...210....1G,2016JCAP...07..022A}), to the smaller scales of dense stellar systems for its consequences on the motion of supermassive black holes (SMBHs) in galactic cores (\citealt{2012ApJ...745...83A,2018ApJ...857L..22T,2020IAUS..351...93D,2021MNRAS.503.6098R,2022MNRAS.510..531C}), globular clusters (GCs) orbiting their host galaxies (\citealt{1989MNRAS.239..549W,1998ApJ...502..150C,2003A&A...405...73B,2006A&A...453....9A,2007A&A...463..921A}), or exotic stellar objects such as for example blue straggler stars (BSS, \citealt{Bellazzini}) in GCs \cite[see][]{Paresce,Bellazzini,Bailyn,     Ferraro2001, Ferraro2009, Ransom, Pooley,2014ApJ...795..169A,2016ApJ...833..252A,Miocchi,Pasquato2018, Pierfrancesco2}.\\
\indent Since the pioneering work of Chandrasekhar, the DF formalism, initially conceived for a point-like particle in an infinitely extended background of scatterers, has been extended to the case of finite-sized objects (e.g. see \citealt{Mulder,2016MNRAS.455.3597Z}) sinking in the host stellar system, flattened or spherical models with self-gravity (e.g. see \citealt{1971ApJ...166..275K,1984MNRAS.209..729T}) or spheroids with anisotropic velocity distribution (\citealt{Binney}).\\ 
\indent More recently, \cite{mond1} and \cite{mond2} derived an expression for the DF formula in the case of modified Newtonian dynamics (hereafter MOND, \citealt{Milgrom,1984ApJ...286....7B}), and performed $N-$body simulations of sinking satellites in MOND, while \cite{Ciotti:20} considered the effect of a mass spectrum for the field particles and \cite{2016JCAP...05..021S} that of a non-thermal, power-law-like velocity distribution.\\
\indent The main conclusion of these works is that, in general, using the original  formulation of the DF for idealized infinite systems (cfr. Eq. \ref{dfchandra}), leads to a substantial underestimation (even of a factor 10) of its effectiveness when applied to more realistic models.\\
\indent Prompted by the detection of gravitational waves (GWs) from binary compact objects announced by the LIGO and VIRGO collaborations \cite[][]{LigoVirgo, 2016LV}, a renewed interest in relativistic stellar dynamics (\citealt{1985ApJ...298...34S,2014MNRAS.443..355H}) has recently widespread, in particular with respect to the formation and migration processes of single and binary black holes (BHs) in GCs (\citealt{2018MNRAS.481.5445S,2018PhRvD..98l3005R,2019MNRAS.486.5008A,2022arXiv220308163T}), Supermassive Black Holes in galactic cores (\citealt{2015ApJ...814...57M,2020PhRvD.102j4002F,2021MNRAS.508.2524K,2022MNRAS.513.4657L}), or runaway objects from dense star clusters (\citealt{2017MNRAS.470.3049R,2020ApJ...900...14F,2022arXiv220401594B}).\\ \indent From the theoretical point of view, if on one hand the distribution function-based approach to relativistic stellar dynamics has been widely explored since \cite{1968ApJ...153..643F} first attempt (see e.g. \citealt{1975ApJ...199..307K,1983AnPhy.150..455E,1984AnPhy.152...30I,1984ApJ...282..361K,1993AnPhy.225..114K,1994PhRvD..49.5115K,2020EPJP..135..290C,2020EPJP..135..310C} and references therein), on the other hand much less has been done in the context of relativistic collisional systems, in the original Chandrasekhar picture. \cite{Lee:69} formally extended Equation (\ref{dfchandra}) accounting for the first post-Newtonian corrections to the gravitational force, though not evaluating it for an explicit choice of the velocity distribution $f(\mathbf{v}_{\rm F})$ or included the effects of relativistic velocities. Nevertheless  he concluded that DF in a dense stellar system, where strong deflections induced by small impact parameters may happen, is always enhanced when considering corrections to the Newtonian force of order $1/c^2$, where $c$ is the speed of light.\\
\indent \cite{Syer:94} derived an alternative expression for the DF in the case of a relativistic test particle crossing an isotropic medium of lighter particles with $m\ll M$ in the limit of small scattering angles. More recently, \cite{2007MNRAS.382..826B}, \cite{2019JCAP...08..017K}, \cite{2021PhRvD.104j3014T}, \cite{2022PhRvD.105h3008V} and \cite{2022PhRvD.105h4041C} explored the onset of DF in relativistic fluids where the test particle induces a wake and \cite{2017PhRvD..95f4014C} included the precession effect induced by the contribution of the so-called Gravitomagnetic effect (e.g. see \citealt{2014GReGr..46.1792C} and references therein).\\
\indent In this work we explore this matter further aiming at formulating a general treatment of DF that can be applied in different regimes dominated by collisional effects, involving very large test particle masses and/or relativistic background systems.\\
\indent The paper is structured as follows: In Section 2 we introduce the astrophysical systems where relativistic collisional processes are relevant. In Section 3 we revise the classical formalism of DF based on the hyperbolic two-body problem, to introduce the fundamental quantities used in the following. In Section 4 we derive the (special) relativistic DF expression, complete with a relativistic velocity distribution. In Section 5 we compute the post-Newtonian $1/c^2$ corrections to the classical DF expression and we show an alternative derivation of the relativistic post-Newtonian DF expression using the Darwin Lagrangian. Finally, Section 6 summarises and sketches the possible applications of this work.
\section{Relativistic stellar dynamical systems}
{When discussing relativistic effects in stellar dynamics (in particular with respect to DF) one can think of three set-ups: {\it i)} The test mass $M$ moves at large speed $v$ so that its relativistic (Lorentz) factor \begin{equation}\label{gammalor}
 \gamma_v=\frac{1}{\sqrt{1-\frac{v^2}{c^2}}}   
\end{equation}
becomes significantly larger than unity. {\it ii)} The test mass $M$ is so large that some degree of post-Newtonian approximation should be used when resolving the close approaches with the background stars. {\it iii)} The velocity distribution of the system $f(v)$ has non negligible relativistic tails.\\
\indent An important example of the first case is represented by the so-called Hypervelocity stars (HVSs). The latter could be produced during multiple strong interactions between the central galactic SMBH and an infalling GC, that can accelerate many stars belonging to the GC to high velocities, even up to eject them in jets from the inner galactic region. This usually may occur as a result of three or four body interactions with the SMBH (or SMBH-binary). Alternatively, dynamical  kicks due to supernovae explosions or close encounters of a hard massive binary star and a single massive star could accelerate stars to extreme velocities. In addition, some galactic HVSs are thought to be the result of a merger event with another nearby galaxy with a high velocity relative to the present galactic environment (see \citealt{high}).\\ 
\indent Typically, HVSs observed in the galactic halo have velocities that can reach up to $\sim 1.2\times10^3$ Km/s whence, the typical stellar velocity in the Galaxy is roughly 100 Km/s. Therefore, the size of the relativistic corrections, quantified in $\gamma-1$, on a HVS are of the order of $8\times 10^{-6}$ while for an average star amount to $5.6\times 10^{-8}$.\\  
\indent The second case is exemplified by a massive black hole moving through a dense stellar system, such as for example a nuclear star cluster (NSCs) or a core-collapsed GC. BHs can be accelerated  themselves to large velocities, for instance during the final stage of SMBHs coalescence by the large recoil due to anisotropic emission of GWs, with $v_{\rm recoil} \approx 10^3$ Km/s for the coalesced object, that may also be displaced from the minimum of the host galaxy potential-well, or even ejected \cite[][]{Marconi, Kim}.\\ 
\indent NSCs have typical mass densities in the range $3\times 10^4 - 2\times 10^5M_{\odot}/{\rm pc}^3$. Assuming an average stellar mass of about $0.5M_{\odot}$ implies that their average inter-star distance $r_{\rm int}$ is of the order of $2\times 10^{-2}$ pc, while the Schwarzschild radius $r_s=2GM_{\rm BH}/c^2$ for a $10^5-10^6M_{\odot}$ massive BH is roughly $10^{-7}-10^{-6}$ pc, and therefore a close encounter with a star at about $10^{-2}r_{\rm int}$ happens at $\sim 10^2r_s$, where relativistic corrections should be taken into account.\\
\indent As an example of the third case, one could think to the distribution of dark matter in proximity of a massive black hole. In general, in the $\Lambda$CDM paradigm, dark matter is supposed to be in a non relativistic and collisionless regime (see e.g. \citealt{BinTre:08}). However, in presence of the deep gravitational potential of a BH, (dark) matter particles could reach relativistic velocities (\citealt{1993ApJ...403..278Z}). Quantifying the relativistic corrections to DF in a dark matter cusp with a central BH, is therefore worth exploring in the context of seeding mechanisms on primordial BHs (\citealt{2020PhRvD.102h3006K,2022arXiv220707576C}).}
%%%%%%%%%%%%%%%%%%%%%%%%%%%%%%%%%%%%%%%%%%%%%%%%%%%%%%%%%%%%%%%%%%%%%%%%%%%%%%%%%%%%%%%
\section{Dynamical friction: the classical case}
Before tackling the problem of its relativistic generalization, it is instructive to revisit the classical treatment of the DF and retrive Equation (\ref{dfchandra}). As usual, we consider each single encounter between the test particle and the field particles as a hyperbolic two body problem in the frame centered on the field particle. Let $(\textbf{x}_{\rm T},\textbf{v}_{\rm T})$ and $(\textbf{x}_{\rm F},\textbf{v}_{\rm F})$ be the positions and velocities of $M$ and $m$, respectively, and let 
\begin{equation}\label{velocità}
\textbf{r}=\textbf{x}_{\rm T}-\textbf{x}_{\rm F};\quad \textbf{V}=\dot{\textbf{r}}= \textbf{v}_{\rm T}-\textbf{v}_{\rm F}
\end{equation}
be their relative position and velocity. We recall that the equation of motion for a fictitious particle of reduced mass $\mu=mM/(M+m)$ moving in the Keplerian potential of the fixed body of mass $M+m$, is
\begin{equation} 
\frac{mM}{m+M}\ddot{\textbf{r}}= -\frac{GMm}{r^{2}}\hat{\textbf{e}}_{\textbf{r}}.
\end{equation}
The energy conservation along the orbit of $\mu$ for a given encounter with impact parameter $b$, implies that the relative velocity vector $\textbf{V}$ is deflected by an angle $\pi-2\psi$, in the orbital plane defined by
\begin{equation} \label{1}
\cos\psi= \frac{1}{ \sqrt{ 1+ \frac{b^{2}V^{4}}{G^2(M+m)^2}}}.
\end{equation}
Using finite differences, we can always express the relative velocity change as
\begin{equation} \label{2}
\Delta\textbf{V}=\Delta\textbf{{v}}_{\rm F}-\Delta\textbf{{v}}_{\rm T},
\end{equation}
where $\Delta\textbf{{v}}_{\rm F}$ and $\Delta\textbf{{v}}_{\rm T}$ are the velocity variations of $m$ and $M$ during the encounter. Since the velocity of the center of mass is constant (by definition) during the encounter, we have that
\begin{equation} \label{3}
m\Delta\textbf{{v}}_{\rm F}+M\Delta\textbf{{v}}_{\rm T}=0.
\end{equation}
Eliminating $\Delta\textbf{{v}}_{\rm F}$ in the two equations above yields
\begin{equation} \label{4} 
\Delta\textbf{{v}}_{\rm T}=-\Bigl(\frac{m}{m+M}\Bigr)\Delta\textbf{V}.
\end{equation}
We must now evaluate $\Delta\textbf{V}$ in order to find $\Delta\textbf{v}_{\rm T}$. The conserved angular momentum per unit mass of the reduced particle is $L=bV$. Let us now label with $\theta_{\rm defl}$ the deflection angle. The relation between the radius and azimuthal angle of a particle on a Keplerian orbit becomes
\begin{equation} \label{5} 
\frac{1}{r}=C\cos(\psi-\psi_{\rm 0})+ \frac{G(M+m)}{b^{2}{V}^2},
\end{equation}
where the constant $C$ and the phase angle $\psi_{\rm 0}=\psi(t=0)$ are determined by the initial conditions.
Deriving (\ref{5}) with respect to the time we obtain 
\begin{equation} \label{6}
\frac{dr}{dt}=Cr^{2}\dot{\psi}\sin(\psi-\psi_{\rm 0})=CbV\sin(\psi-\psi_{\rm 0}),
\end{equation}
where the last term arises from $L=r^{2}\dot{\psi}$. If we impose that $\psi=0$ when $t\rightarrow -\infty$ we obtain from (\ref{6})
\begin{equation}
-V=Cb\sin(-\psi_{\rm 0}).
\end{equation}
Evaluating equation (\ref{5}) we then have 
\begin{equation}
0=C\cos(\psi_{\rm 0})+ \frac{G(M+m)}{b^2 V^2},
\end{equation}
and eliminating $C$ from the equations above we obtain 
\begin{equation}\label{psi0}
\tan\psi_{\rm 0}=-\frac{b V^2}{G(M+m)}.
\end{equation}
From Equations (\ref{5}) and (\ref{6}) we can appreciate that the point of closest approach is reached when $\psi=\psi_{\rm 0}$ and, since the orbit is symmetrical about this point, the deflection angle is $\theta_{\rm defl}=2\psi_{\rm 0}-\pi$.
Thanks to the conservation of energy, after the encounter, the modulus of the relative velocity, equals the modulus of the initial relative velocity and therefore, the components of $\Delta\textbf{V}$, parallel and perpendicular to the initial relative velocity vector $\textbf{V}$, $\Delta\textbf{V}_{\rm \parallel}$ and $\Delta\textbf{V}_{\perp}$ become
\begin{equation}\label{7}
||\Delta\textbf{V}_{\perp}||=\frac{2bV^3}{G(M+m)}{\Biggl[1+\frac{b^2 V^4}{G^{2}(M+m)^{2}}}\Biggr]^{-1};
\end{equation}
and
\begin{equation}\label{8}
||\Delta\textbf{V}_{\rm \parallel}||=2V\Biggl[1+\frac{b^2 V^4}{G^{2}(M+m)^{2}}\Biggr]^{-1}.
\end{equation}
In a homogeneous background of particles equal masses $m$, all $\Delta\textbf{v}_{\rm T\perp}$ sum to zero by symmetry, (using the so-called "Jeans swindle", see e.g. \citealt{BinTre:87}), while the parallel velocity changes add up, and thus the mass $M$ will experience a deceleration (cfr Eq. (\ref{4})) as a result of the DF. Therefore, it is sufficient to evaluate $\Delta\textbf{v}_{\rm T\parallel}$ as
\begin{equation}\label{9}
||\Delta\textbf{v}_{\rm T\parallel}||=\frac{2mV}{M+m}\Biggl[1+\frac{b^2 V^4}{G^2 (M+m)^2}\Biggr]^{-1}.
\end{equation}
In a system defined by the phase-space distribution function $F=nf(\textbf{v}_{\rm F})$, where $n$ is a constant number density and $f(\textbf{v}_{\rm F})$ be the velocity distribution, the rate at which the mass $M$ encounters stars with impact parameter between $b$ and $b + db$, and velocities between $\textbf{v}_{\rm F}$ and $\textbf{v}_{\rm F}+d\textbf{v}_{\rm F}$, is 
\begin{equation}\label{nenc}
n_{\rm enc}=2\pi bdb V n f(\textbf{v}_{\rm F}) {d^{3}}\textbf{v}_{\rm F},
\end{equation}
where $d^3\textbf{v}_{\rm F}$ is the velocity-space element. The total change in velocity suffered by $M$ is found by adding all the contributions of $||\Delta\textbf{v}_{\rm T\parallel}||$ due to particles with impact parameters from $b_{\rm min}$ to a $b_{\rm max}$ and then summing over all velocities of stars as
%\begin{strip}
\begin{multline}\label{10}
\begin{gathered}
\frac{d\textbf{v}_{\rm T}}{dt}\bigg|_{\textbf{v}_{\rm F}}=\textbf{V} n f(\textbf{v}_{\rm F}) {d^{3}}\textbf{v}_{\rm F} \int_{b_{\rm min}}^{b_{\rm max}} {||\Delta\textbf{v}_{\rm T\parallel}|| 2\pi b db}=\\
=\textbf{V} n f(\textbf{v}_{\rm F}) {d^{3}}\textbf{v}_{\rm F}\int_{b_{\rm min}}^{b_{\rm max}}{\frac{2mV}{M+m}\Biggl[1+\frac{b^2 V^4}{G^2(M+m)^2}\Biggr]^{-1} 2\pi b db}.
\end{gathered}
\end{multline}
%\end{strip}
Let us first perform the integral over $b$
\begin{multline}\label{12}
\begin{gathered}
\int_{b_{\rm min}}^{b_{\rm max}}{\frac{2m V}{M+m}\Biggl[1+\frac{b^2V^4}{G^{2}(M+m)^2}\Biggr]^{-1} 2\pi b db}=\\
=\frac{2\pi G^{2}(M+m)}{V^3}\log{   {\Biggl[\frac{1+\frac{{b_{\rm max}^{2}V^4}}{G^{2}(M+m)^2} }{1+\frac{{b_{\rm min}^{2}}V^{4}}{G^{2}(M+m)^2}}}\Biggr]}.
\end{gathered}
\end{multline}
We must note that, choosing the minimum and maximum impact parameters is a rather delicate step. When using the impulsive approximation (i.e. $\mu||\Delta\mathbf{V}_\perp||=2GMm/bV$; see e.g. \citealt{Ciotti:20}), the so-called "ultraviolet divergence" occurs when performing the integral over $b$ in (\ref{10}) and setting $b_{\rm min}=0$. The latter divergence is actually artificial as it disappears when the full solution of the hyperbolic two body problem is taken into account as in this discussion. Conversely, the "infrared divergence" appearing for $b\rightarrow\infty$ cannot be eliminated in an infinite system and therefore one must put a upper cutoff with a suitable choice of $b_{\rm max}$.\\
\indent Not surprisingly, this point is still source of debate (e.g. see the discussion in \citealt{VanAlbada} and references therein). The problem lies in what should dominate between few strong encounters with nearby stars (see \citealt{Chandra:41, Chandra42, Chandra:43}, see also \citealt{Kandrup}), or many weak encounters with distant stars (see \citealt{Spitzer,BinTre:87}). In the first interpretation $b_{\rm max}$ should be of the order of the average inter-particle distance, while in the second, it should be of the order of size of the system. Both views are in principle plausible,
the former as it is more intuitive to think that the largest contribution must be due to nearest stars and the latter because there is no screening in gravitational systems, at variance with (quasi-)neutral Plasmas where charges of opposite signs are present. In this work we follow the Spitzer approach.\\
\indent We note that, under most conditions of practical interest in astrophysics, the quantity $\Lambda^2={b^{2}}V^{4}/G^{2}(M+m)^2$ is typically much greater than unity. 
For this reason, we will now on replace $\log 1+\Lambda^2$ with $2\log\Lambda$, 
recovering the widely used definition of the Coulomb logarithm as $\log{  \left[{{b_{\rm max}^{2}}V^{4}}/{G^{2}(M+m)^2}   \right]    }$\footnote{In most systems of astrophysical interest $\log\Lambda$ is a number of order 10. For example, in a globular cluster of mass $M_{\rm GC}=10^{6} M_{\odot}$ sinking at $V\sim 100 \, \rm{km/s}$ through a galaxy of radius $b_{\rm max}\approx 2 \, \rm{kpc}$ in which the stars have mass of the order of one Solar mass, we have that $\log{  \left[  1+ {{b_{\rm max}^{2}}V^{4}}/{G^{2}(M+m)^2}   \right]    }\approx 17$}.\\
\indent In this approximation, the right hand side of Eq. (\ref{12}) becomes 
\begin{gather} \label{11}
\frac{  2 \pi G^{2}(M+m) }{{V^{3}}}\log{   {\Biggl[   \frac { 1+ \frac{{b_{\rm max}^{2}}V^{4}}{G^{2}(M+m)^2} }{  1+ \frac{{b_{\rm min}^{2}}V^{4}}{G^{2}(M+m)^2  } }   }\Biggr]} \approx \frac{  4 \pi G^{2}(M+m) }{V^{3}}\log{\Lambda}.
\end{gather}
Combining the expression above with Eq. (\ref{10}) yields 
\begin{equation} \label{13}
\frac{d\textbf{v}_{\rm T}}{dt}=-4\pi G^{2}nm(M+m) \log{\Lambda}\int{f(\textbf{v}_{\rm F})\frac{ (\textbf{v}_{\rm T}-\textbf{v}_{\rm F}) }{||\textbf{v}_{\rm T}-\textbf{v}_{\rm F}||^{3}}d^{3}\textbf{v}_{\rm F}  },
\end{equation}
where we have replaced $\textbf{V}$ with its definition (cfr. Eq. \ref{velocità}) and assumed $\log\Lambda$ as the {\it velocity averaged} Coulomb logarithm (e.g. see the discussion in \citealt{2021isd..book.....C}). The velocity integral in Eq. (\ref{13}) is often referred to as the first Rosenbluth potential (see e.g. \citealt{1957PhRv..107....1R}).\\
\indent Remarkably, the problem of computing the acceleration ${d\textbf{v}_{\rm T}}/{dt}$ integrating over all field star velocities, is formally equivalent to that of evaluating the gravitational field at  $\textbf{v}_{\rm T}$ generated by the "mass density" $\rho(\textbf{v}_{\rm F})=4\pi \log{ \Lambda}Gm(M+m)f(\textbf{v}_{\rm F})$. Assuming an isotropic (spherically symmetric) velocity distribution, in virtue of the second Newton's theorem (e.g. see \citealt{Chandra:95}) we have that only the stars such that $v_{\rm F}<v_{\rm T}$ contribute to the slowing down of $M$, hence
\begin{equation}\label{16}
\frac{d\textbf{v}_{\rm T}}{dt}=-16 {\pi}^{2} G^{2}nm(M+m) \log{\Lambda} \frac{\textbf{v}_{\rm T}}{{v^3_{\rm T}}}  \int_{0}^{v_{\rm T}} {f(v_{\rm F}){v^2_{\rm F}} dv_{\rm F}}.
\end{equation}
In the special case where $f(v_{\rm F})$ is a Maxwellian with dispersion $\sigma$, 
\begin{equation}\label{MBdistribution}
f(v_{\rm F})= {  \frac{v_{\rm F}^2}{(2\pi {\sigma}^{2})^{3/2}}}{\exp{\biggl(-\frac{v_{\rm F}^{2}}{2 {\sigma^{2}}}}\biggr)},  
\end{equation}
evaluating the velocity volume function integral in Eq. (\ref{16}) yields
\begin{equation}\label{17}
\frac{d\textbf{v}_{\rm T}}{dt}=-4\pi G^2 (M+m)\rho\log\Lambda\left[ \Erf  \left(\frac{v_{\rm T}}{\sqrt{2}\sigma}\right)  - \frac{2 v_{\rm T} e^{-\frac{v_{\rm T}^2}{2\sigma^2}}}{\sqrt{2\pi}\sigma} \right]\frac{\textbf{v}_{\rm T}}{v_{\rm T}^3},
\end{equation} 
where we have condensed $nm$ in $\rho$ (the mean mass density of the field particles), and $\Erf(x)$ is the standard error function defined (see \citealt{Weber}) as
\begin{equation}
\Erf(x)= \frac{2}{\sqrt{\pi}}\int_{0}^{x}e^{-t^2}dt.
\end{equation}
\section{Dynamical friction: the special relativistic generalization}\label{relativisticDF}
As mentioned above, the Chandrasekhar DF formula was extended to relativistic velocities by \cite{Syer:94}, but only in the weak scattering limit (i.e. $b\gg r_{\rm s}\approx {GM}/{c^2}$ and small deflection angle $\theta_{\rm defl}$). We will now derive a more general expression for a generic $\theta_{\rm defl}$, therefore also accounting for the case of strong scattering, (i. e. $\theta_{\rm defl} \rightarrow \pi/2$).\\
\indent In this derivation we will keep the classical $1/r^2$ Newtonian force and replace velocity composition with its relativistic counterpart. For this purpose, we define an inertial frame $\textit{S}^\prime$, in which the test star of mass $M$ is stationary at the beginning of the encounter (i. e. the field star $m$ is at infinity). In said frame, $\theta_{\rm defl}=\pi-2\psi$, (with $\psi$ given by Eq \ref{1}), is the deflection angle according to the classical unbound two body problem.  Such assumption is motivated by the fact that in the astrophysically relevant case where $M>m$, the relativistic scattering angle in an elastic collision has the same majorant as the classical case (see \citealt{Landau}, Chapter 2) given by $\sin\theta_{\rm defl,max}=m/M$ (see also \citealt{alma991014033969704336}). In this approximation, however, we are not assuming a small angle (i.e. weak scattering) limit as no assumption has been made on the relative angles between $\mathbf{v}_{\rm T}$ and $\mathbf{v}_{\rm F}$.\\
\indent In the (special) relativistic encounter the relative velocity $V$ becomes (e.g. see \citealt{Landau})
\begin{equation}\label{vrelrel}
V^2=\frac{{||\textbf{v}_{\rm T} - \textbf{v}_{\rm F}||^2  -  \frac{1}{c^2} {||\textbf{v}_{\rm T}\wedge\textbf{v}_{\rm F}||^2 }}   }{\bigl( 1- \frac{   {\textbf{v}_{\rm T} }\cdot {\textbf{v}_{\rm F}  } }{   c^2   }    \bigr)^2  },
\end{equation} 
for arbitrary choices of $\textbf{v}_{\rm T}$ and $\textbf{v}_{\rm F}$. We stress the fact that, in the relativistic case, the velocity $\textbf{v}_{\rm A|B}$ of a body $A$ respect to another $B$ is {\it not} equal to $-\textbf{v}_{\rm B|A}$ of $B$ with respect to $A$. This loss of symmetry is related to the \cite{doi:10.1080/14786440108564170} precession\footnote{In practice, two subsequent non-collinear boosts are equivalent to the composition of a rotation of the coordinate system and a boost.} and the fact that two subsequent Lorentz transformations rotate the coordinate system (cfr. \citealt{Weinberg}). Said rotation however, has conveniently no effect on the magnitude of a vector and hence, the modulus of the relative velocity is symmetrical.\\
\indent Let $\mathbf{v}_{\rm F,\mu}=\gamma_{v_{\rm F}}(c, \textbf{v}_{\rm F})$ and $\mathbf{v}_{\rm T,\mu}=\gamma_{v_{\rm T}}(c, \textbf{v}_{\rm T})$ be the four-velocities of the field and test stars, respectively  and where $\gamma_{v_{\rm F;T}}=\big({1-{v_{\rm F;T}^2}/{c^2}}\big)^{-1/2}$ are the Lorentz factors of $M$ and $m$. Let us consider a single encounter in the "laboratory frame" $\textit{S}$. This process can be expressed as a product of a Lorentz boost $\Gamma$ in $\textit{S}^\prime$, a rotation $\mathcal{R}(\theta_{\rm defl})$ in the 3-space and, finally, an inverse boost $\Gamma^{-1}$, reverting back to $S$.\\
\indent Defining $\mathbf{p}_{{\rm F}\mu}=m\gamma_{v_{\rm F}}(c, \textbf{v}_{\rm F})$ as the 4-momentum of $m$ before the encounter, we have, that ${\mathbf{p}_{\rm F}^\prime}^{\mu}= \Lambda^{\mu}_{\,\,\,\, \nu} \, \mathbf{p}_{\rm F}^{\nu}$, where $\Lambda=\Gamma^{-1} \mathcal{R}(\theta_{\rm defl})\Gamma$; and thus we formally obtain
\begin{equation}
\Delta \mathbf{p}_{\rm F}^{\mu}= (\Gamma^{-1} \mathcal{R}(\theta_{\rm defl})\Gamma -1) \mathbf{p}_{\rm F}^{\mu}.
\end{equation}
Since the motion is planar, as we are still dealing with a classical two body problem, we can simplify the notation involving 4-vectors by using 3-vectors instead, where only two of space dimensions are maintained; one parallel and one perpendicular to $\textbf{v}_{\rm T}$. However, as argued before, by reasons of symmetry, only the parallel component of $V$ contributes to the DF. Denoting with $\phi$, the angle between $\textbf{v}_{\rm T}$ and $\textbf{v}_{\rm F}$, we can now write
\begin{equation}
\mathbf{p}_{\rm F}^{\mu}=
m\gamma_{v_{\rm F}}
\begin{pmatrix} c\\
v_{\rm F}\cos\phi \\
v_{\rm F}\sin\phi \end{pmatrix}. 
\end{equation}
With such choice $\Gamma$, $\mathcal{R}$ and $\Gamma^{-1}$ read
\begin{multline}
\begin{gathered}\label{gammathetagamma}
\Gamma=
\begin{pmatrix} \gamma_{v_{\rm T}} & -\frac{v_{\rm T}}{c} \gamma_{v_{\rm T}} & 0\\
 -\frac{v_{\rm T}}{c} \gamma_{v_{\rm T}} & \gamma_{v_{\rm T}} & 0 \\
 0 & 0 & 1
\end{pmatrix}, \,\, 
\mathcal{R}(\theta_{\rm defl})=
\begin{pmatrix} 1 & 0 & 0\\
 0 & \cos\theta_{\rm defl} & -\sin\theta_{\rm defl} \\
 0 & \sin\theta_{\rm defl} & \cos\theta_{\rm defl}
 \end{pmatrix}, \\
 \Gamma^{-1}=
 \begin{pmatrix} \gamma_{v_{\rm T}} & \frac{v_{\rm T}}{c} \gamma_{v_{\rm T}} & 0\\
 \frac{v_{\rm T}}{c} \gamma_{v_{\rm T}} & \gamma_{v_{\rm T}} & 0 \\
 0 & 0 & 1
 \end{pmatrix}.
 \end{gathered}
 \end{multline}
After the encounter the 4-momentum of the field star $\mathbf{p}_{\rm F}^{\mu}$ changes by $\Delta \mathbf{p}_{\rm F}^{\mu}$ defined as
 \begin{equation}\label{190}
 \Delta \mathbf{p}_{\rm F}^{\mu}= m\gamma_{v_{\rm F}} 
 \begin{pmatrix} A\\
 B\\
 C
\end{pmatrix} 
\end{equation}
where, respectively,
\begin{multline}\label{abc}
\begin{gathered}
A=c\gamma^2_{v_{\rm T}}-\frac{v_{\rm T}v_{\rm F}}{c}\gamma^2_{v_{\rm T}} \cos\phi+ \frac{v_{\rm T}}{c} \gamma_{v_{\rm T}} [\gamma_{v_{\rm T}} (v_{\rm F}\cos\phi-v_{\rm T})\,\times\\
\times\,\cos\theta_{\rm defl}-v_{\rm F}\sin\phi \sin\theta_{\rm defl} ]-c \\
B=v_{\rm T}\gamma^2_{v_{\rm T}}\bigg(1 - \frac{v_{\rm T}v_{\rm F}}{c^2}\cos\phi\bigg) + \gamma_{v_{\rm T}}  [ \gamma_{v_{\rm T}}(v_{\rm F}\cos\phi-v_{\rm T})\,\times\\ 
\times\,\cos\theta_{\rm defl}-v_{\rm F}\sin\phi \sin\theta_{\rm defl} ] -v_{\rm F}\cos\phi\\
C=\gamma_{v_{\rm T}}( v_{\rm F}\cos\phi-v_{\rm T})\sin\theta_{\rm defl} +v_{\rm F}\sin\phi\cos\theta_{\rm defl}-v_{\rm F}\sin\phi,
\end{gathered}
\end{multline}
and where 
\begin{equation} \label{120}
\theta_{\rm defl}= \pi-2\cos^{-1} \left( \frac{1}{\sqrt{1+\frac{b^2 V^4}{G^2(M+m)^2}}}\right) 
\end{equation}
is the deflection angle. We now have to multiply $\Delta \mathbf{p}_{\rm F}^{\mu}$ for the differential number of encounters $dn^\prime_{\rm enc}= 2\pi n\mathcal{V}d^3\textbf{v}_{\rm F} bdbdt$ in the laboratory frame. The latter, at variance with the one given by Eq. (\ref{nenc}), is a Lorentz-invariant quantity\footnote{Notably, due to the relativistic length contraction, the number density $n$ along the direction of $M$ would increase in the rest frame of $m$, cfr. \cite{Landau}.}, where
\begin{equation}
\mathcal{V}\equiv V \big(1-{\textbf{v}_{\rm T}}\cdot{\textbf{v}_{\rm F}}/c^2 \big)=V\big(1-v_{\rm T}v_{\rm F}\cos\phi/c^2\big).
\end{equation}
In integral form, the momentum variation of the field particle $m$ is now given by
\begin{equation} \label{18}
\frac{d\mathbf{p}_{\rm F}^{\mu}}{dt}=2\pi n \int\int{ b  \Delta \mathbf{p}_{\rm F}^{\mu} \mathcal{V} f(\textbf{v}_{\rm F})dbd^3\textbf{v}_{\rm F}  }.
\end{equation}
The momentum of the test particle $M$, $p_{\rm T}^{\mu}= M\gamma_{v_{\rm T}}(c, \textbf{v}_{\rm T})$, will suffer the opposite change
\begin{equation} \label{19}
\frac{d\mathbf{p}_{\rm T}^{\mu}}{dt}=-\frac{d\mathbf{p}_{\rm F}^{\mu}}{dt}.
\end{equation}
To evaluate  Equation (\ref{18}) we need first to substitute the expression for the cosine and sine of the  deflection angle (\ref{120}) in Eqs. (\ref{gammathetagamma}-\ref{abc}) that read
\begin{equation}\label{sincos}
\cos{\theta_{\rm defl}}= \frac{\frac{b^2V^4}{G^2(M+m)^2}-1}{\frac{b^2V^4}{G^2(M+m)^2}+1};\quad \sin{\theta_{\rm defl}}=\frac{ \frac{2bV^2}{G(M+m)}   }{1+\frac{b^2V^4}{G^2(M+m)^2 }}.
\end{equation}
We stress the fact that for the derivation the DF formula, only the parallel component of $\textbf{v}_{\rm T}$ contributes, so it is sufficient to evaluate the following expression for the parallel component $p_{\rm F \parallel}$:
\begin{multline}\label{pfpara}
\begin{gathered} 
\frac{dp_{\rm F\parallel}}{dt}=2\pi n m\int\int{ b} \gamma_{v_{\rm F}} \mathcal{V} f(\textbf{v}_{\rm F})\Bigg\{  v_{\rm T}\gamma^2_{v_{\rm T}}\bigg(1 - \frac{v_{\rm T}v_{\rm F}}{c^2}\cos\phi\bigg)+\\
+\gamma_{v_{\rm T}}\bigg[ \gamma_{v_{\rm T}} (v_{\rm F}\cos\phi-v_{\rm T})\frac{\frac{b^2V^4}{G^2(M+m)^2} -1 }{   \frac{b^2V^4}{G^2(M+m)^2}+1}
-v_{\rm F}\sin\phi \times \\ 
\times\frac{ \frac{2bV^2}{G(M+m)}   }{\frac{b^2V^4}{G^2(M+m)^2 }+1}\bigg]-v_{\rm F}\cos\phi \Bigg\}db {d^3} 
\textbf{v}_{\rm F}.
\end{gathered}
\end{multline}
Following the classical derivation discussed in Sect. 3, we perform first the integral over the impact parameter $b$. As we are considering an isotropic $f(\mathbf{v}_{\rm F})$, we assume $\phi$ such that all odd terms involving $\sin\phi$ zero-out. We then apply Eq. (\ref{19}) obtaining the deceleration on the momentum of particle $M$ and then we obtain the DF formula for the velocity $\mathbf{v}_{\rm T}$ one dividing by $\gamma_{v_{\rm T}}M$ as
\begin{multline}\label{integ}
\begin{gathered}
\frac{d\mathbf{v}_{\rm T}}{dt}=-4\pi G^2\frac{(M+m)^2}{M}\rho \gamma_{v_{\rm T}}\log\Lambda \int {\frac{\gamma_{v_{\rm F}}\mathcal{V}f(\textbf{v}_{\rm F})(\mathbf{v}_{\rm T}-\mathbf{v}_{\rm F})}{V^4}}d^3\textbf{v}_{\rm F}\\
\end{gathered}
\end{multline}
where, again, we made use of the limit  $\log(1+\Lambda^2)\rightarrow\log\Lambda$ for large values of $\Lambda$ and substituted $\rho$ to $mn$.\\
\indent As we are accounting for relativistic velocities and transformations, when performing the velocity integral in (\ref{integ}), with $V$ given by Eq. (\ref{vrelrel}), we need to use a velocity distribution function in covariant form. For example, when considering a thermalized relativistic gas, a natural choice is the Maxwell-J{\"u}ttner distribution (see \citealt{juttner}, see also \citealt{1968ApJ...153..643F})
\begin{equation}\label{juttner}
f(v_{\rm F})= \frac{\gamma_{v_{\rm F}}^5v_{\rm F}^2}{  c^3\Theta \, \mathcal{K}_{\rm 2}(\Theta^{-1})} \exp \bigg(- \frac{\gamma_{v_{\rm F}}}{\Theta}\bigg).
\end{equation}
%%%%%%%%%%%%%%%%%%%%%%%%%%%%%%%%%%%%%%%%%%%%%%%%%%%%%%%%%%%%%%%%%%%%%%%%%%%%%%%%%%%%%%%
\begin{figure*}
\includegraphics[width = 0.95\textwidth]{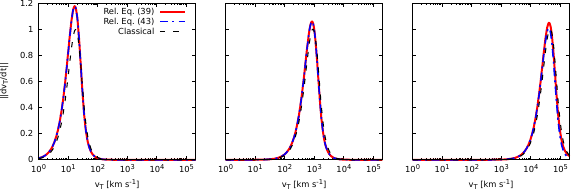}
\caption{Relativistic (Eq. 39 red solid lines, and Eq. 43 blue dotted dashed lines) and classical (dashed lines) dynamical friction force as function of the test particle velocity $v_{\rm T}$ for three different cases with isotropic relativistic distribution: A 10 $M_{\odot}$ BH in a star cluster with velocity dispersion $\sigma\sim10$ Km ${\rm s}^{-1}$ and core density of  $10^2M_{\odot}{\rm pc}^{-3}$ (left panel). A $10^7$ $M_{\odot}$ BH in a galactic core with $\sigma\sim 500$ Km ${\rm s}^{-1}$ and density of $4\times 10^6M_{\odot}{\rm pc}^{-3}$ (mid panel). A stellar mass BH in a relativistic dark matter cusp around a SMBH with $\sigma\sim 2\times 10^4$ Km ${\rm s}^{-1}$ and mean density of $7\times 10^{t}$ g cm$^{-1}$ (right panel). In all cases the relativstic and classical curves are normalized to the value of main peak of the latter.}
\label{spec}
\end{figure*}
%%%%%%%%%%%%%%%%%%%%%%%%%%%%%%%%%%%%%%%%%%%%%%%%%%%%%%%%%%%%%%%%%%%%%%%%%%%%%%%%%%%%%%
In the expression above $\Theta=\sigma^2/3c^2$ and $\mathcal{K}_{\rm 2}(x)$ is the modified Bessel function of the second kind (see e.g. \citealt{Weber}), often somewhat improperly dubbed Neumann function.\\
\indent In the limit of $c\rightarrow +\infty$ Equation (\ref{integ}) becomes Equation (\ref{13}) with $(M+m)^2/M$ in lieu of $(M+m)$. The reason of this discrepancy with the classical case is due to the fact that, in the relativistic treatment we used the total {\it relativistic} momentum conservation of Equation (\ref{19}), whereas classically one uses Eqs. (\ref{2}-\ref{4}). The application of the latter would in principle commute with the limit $c\rightarrow +\infty$ (and the term $(M+m)$ at the denominator in \ref{4} would elide with another identical factor in \ref{integ}), while Eq. (\ref{19}) loses meaning in such limit, as the time component would diverge.\\
\indent We note that \cite{1972ASSL...31...13K} also encounters a similar disagreement between the Chandrasekhar expression and the particle limit of his fluid formalism, i.e. the frictional force on the test mass is proportional to $M^2m$ rather than $Mm(M+m)$. However, the reason of this difference is ascribed to the fluid picture where the force on $M$ is due to a continuous mass density distribution (see e.g. \citealt{Kandrup}).\\
\indent We note also that, in the fully kinetic approach on DF based on the Fluctuation-Dissipation theorem, pioneered among the others by \cite{1992ApJ...390...79B}, one recovers the original expression formulated by Chandrasekhar with the term $Mm(M+m)$.\\
\indent This established, we estimated the relativistic DF for three cases of astrophysical interest. A 10 $M_{\odot}$ BH in a star cluster with velocity dispersion $\sigma\sim10$ Km ${\rm s}^{-1}$ and core density of  $10^2M_{\odot}{\rm pc}^{-3}$; a $10^7$ $M_{\odot}$ BH in a galactic core with $\sigma\sim 500$ Km ${\rm s}^{-1}$ and density of $4\times 10^6M_{\odot}{\rm pc}^{-3}$ and, finally, stellar mass BH in a relativistic dark matter cusp around a SMBH with $\sigma\sim 2\times 10^4$ Km ${\rm s}^{-1}$ and mean density of $7\times 10^{3}$ g cm$^{-1}$. We always assumed a isotropic relativistic velocity distribution as given in Eq. (\ref{juttner}) when solving (numerically) the integral in Eq. (\ref{integ}). In Figure \ref{spec} (red solid lines) we compare it with the classical expression (Eq. \ref{dfchandra}, black dashed lines). In agreement with \cite{Syer:94} we observe for all systems that, even if $v<\sigma$ the relativistic DF is augmented with respect to its classical counterpart, up to a factor 1.1, that is in general much smaller than the $16/3\gamma_{v_{\rm T}}$ found by Syer. For $v\gtrsim\sigma$ the relativistic DF force is slightly smaller than the classical force while at large $v_{\rm T}$ increases again due to the diverging prefactor $\gamma_{v_{\rm T}}$ in the limit $v_{\rm T}\rightarrow c$.\\
\indent We note that, for relativistic power-law velocity distributions, Equation (\ref{integ}) would become for a given $v_{\rm T}$ considerably larger than (\ref{dfchandra}), due to the contribution of large $v$ tails. We speculate that this could be relevant in the context of plasma physics where (multiple) power-law velocity distributions are often encountered (see e.g. \citealt{doi:10.1063/1.2219428}). In fact, the derivation of Equation (\ref{integ}) can be carried out in a similar fashion for a charge $q_{\rm T}$ deflected in a plasma, as the impact parameter and velocity integrals are the same, the only difference being the dimensional factor containing the masses and the gravitational constant. We note also that Eq. (\ref{integ}) has been obtained in the assumption that the classical angle $\psi_0$ given in Equation (\ref{psi0}) holds even in the case of relativistic velocities. To be more rigorous, one should use instead its relativistic generalization given by
\begin{equation}\label{psi0rel}
\tan\tilde{\psi}_{\rm 0}=-\frac{L\sqrt{2E}}{G(M+m)},
\end{equation}
where $L=\gamma_{V}Vb$ and $E=c^2(\gamma_V-1)$ are the norm of the spatial part of the specific relativistic angular momentum and the specific relativistic kinetic energy, respectively, and $\gamma_V$ is the Lorentz factor of the relative velocity $V$.\\
\indent With such a choice, $\tan{\psi}_{\rm 0}$ and $\tan\tilde{\psi}_{\rm 0}$ differ by the multiplicative factor $\gamma_V\sqrt{2(\gamma_V-1)}c/V$. The latter increases significantly the value of $\tan\psi_0$ only for $V\gg 0.5c$. In this limit, expanding the square roots arising from $\gamma_V$, the terms containing $b^2V^4/G^2(M+m)^2$ in Equations (\ref{120}-\ref{sincos}) in the derivation above, are augmented by a factor $(1+V^2/2c^2)^2$, so that Eq. (\ref{pfpara}) becomes
\begin{multline}\label{pfpara2}
\begin{gathered} 
\frac{dp_{\rm F\parallel}}{dt}=2\pi n m\int\int{ b} \gamma_{v_{\rm F}} \mathcal{V} f(\textbf{v}_{\rm F})\Bigg\{  v_{\rm T}\gamma^2_{v_{\rm T}}\bigg(1 - \frac{v_{\rm T}v_{\rm F}}{c^2}\cos\phi\bigg)+\\
+\gamma_{v_{\rm T}}\bigg[ \gamma_{v_{\rm T}} (v_{\rm F}\cos\phi-v_{\rm T})\frac{\frac{b^2V^4}{G^2(M+m)^2} -\big(1+\frac{V^2}{2c^2}\big)^{-2} }{   \frac{b^2V^4}{G^2(M+m)^2}+\big(1+\frac{V^2}{2c^2}\big)^{-2}}
-v_{\rm F}\sin\phi \times \\ 
\times\frac{ \frac{2bV^2}{G(M+m)}\big(1+\frac{V^2}{2c^2}\big)  }{\frac{b^2V^4}{G^2(M+m)^2 }\big(1+\frac{V^2}{2c^2}\big)^{2}+1}\bigg]-v_{\rm F}\cos\phi \Bigg\}db {d^3} 
\textbf{v}_{\rm F}.
\end{gathered}
\end{multline}
The integration over the impact parameter $b$ can be carried out first in the same fashion as above, yielding a velocity averaged Coulomb logarithm now multiplied by the prefactor $(1+V^2/2c^2)^{-2}$, so that Eq. (\ref{integ}) becomes
\begin{multline}\label{integbis}
\begin{gathered}
\frac{d\mathbf{v}_{\rm T}}{dt}=-4\pi G^2\frac{(M+m)^2}{M}\rho \gamma_{v_{\rm T}}\log\Lambda \int {\frac{\gamma_{v_{\rm F}}\mathcal{V}f(\textbf{v}_{\rm F})(\mathbf{v}_{\rm T}-\mathbf{v}_{\rm F})}{\big(1+\frac{V^2}{c^2}\big)^2V^4}}d^3\textbf{v}_{\rm F}.
\end{gathered}
\end{multline}
Equation (\ref{integbis}) differs significantly from the simplified expression (\ref{integ}) only in the limit of large velocity dispersion (and for large velocities), as shown by the blue dotted-dashed lines in the right panel of Fig \ref{spec}.
%%%%%%%%%%%%%%%%%%%%%%%%%%%%%%%%%%%%%%%%%%%%%%%%%%%%%%%%%%%%%%%%%%%%%%%%%%%%%%%%%%%%%%%%%%%%%%%%%%%%%
\section{Post-Newtonian approximation}
So far, we have derived a formal generalization of the dynamical friction formula in the limit of large test particle velocities or relativistic velocity distributions assuming classical Newtonian forces. We now carry out an alternative derivation involving strong gravitational scattering in the post-Newtonian regime.\\
\indent Introduced by \cite{Einstein} (see also \citealt{Weinberg,Blanchet} and references therein) to study the precession of the perihelion of Mercury, the post-Newtonian approximation consists in an expansion in orders of the parameter $v/c$, such that at the zero-th order it reduces to Newtonian gravity, while at higher orders ($n$PN) the acceleration on the mass $m$ due to the mass $M$ is augmented by corrections of order $\big({v}/{c}\big)^{2n}\sim\big({GM}/{rc^2}\big)^{n}$.
\subsection{Non relativistic velocities}
The cores of dense star clusters are often dominated by massive objects due to dynamical mass segregation, it is therefore interesting to evaluate the DF on a test particle in such an environment where strong scattering by large masses are likely to happen, even though the velocity distribution might not be relativistic. To do so, we begin with a naive derivation of ${d\mathbf{v}_{\rm T \parallel}}/{dt}$ in impulsive approximation keeping the Galilean transformations of velocities, but using the $1$PN-acceleration
\begin{multline}\label{postn}
\textbf{a}_{\rm 1PN}= -\frac{G(M+m)}{r^2} \frac{\textbf{r}}{r}+\frac{G(M+m)}{c^2r^2}\bigg\{\bigg[(4+2\eta)\frac{G(M+m)}{r}+\\
-(1+3\eta)V^2 +\frac{3}{2}\eta \,{\dot{\textbf{r}}}^2 \bigg]\frac{\textbf{r}}{r} +(4-2\eta)\dot{\textbf{r}}\,\textbf{V} \bigg\},
\end{multline}
in the frame centered on the field particle $m$ (see \citealt{2004PhRvD..69j4021M,2017PhRvD..95f4014C}). In the equation above $\mu$, $\mathbf{V}$ and $\mathbf{r}$ have the same meaning as in Sect. 2 and $\eta ={\mu}/(M+m)$. In impulsive approximation we can express the velocity change of the test particles in a discrete time interval $\Delta t=2b/V$ as $||\Delta V_{\rm T}|| \sim a\Delta t=2ab/V$, so that, see e.g. \cite{2021isd..book.....C}, its parallel component becomes
\begin{equation}
{\Delta \mathbf{v}_{\rm T \parallel}} \sim -\frac{\mu}{M}\frac{||\Delta V_{\rm \perp}||^2}{2V^2} \textbf{V}.
\end{equation}
Defining $\textbf{r}= r\hat{\textbf{r}} \sim b \hat{\textbf{r}}$ and $\textbf{V}= V\,\hat{\textbf{V}}$, we obtain the perpendicular relative velocity change as
\begin{multline}
\Delta \textbf{V}_{\perp}\sim \textbf{a}_{\rm 1PN} \frac{2b}{V}\sim -\frac{2G(M+m)}{Vb}  \hat{\textbf{r}} + \frac{2G(M+m)}{c^2Vb}\bigg\{\bigg[(4+2\eta)\frac{G(M+m)}{b}+\\
-(1+3\eta)V^2+\frac{3}{2}\eta V^2\bigg]\hat{\textbf{r}}+(4-2\eta)V^2 \hat{\textbf{V}} \bigg\},
\end{multline}
and its square as
\begin{multline}
||\Delta \textbf{V}_{\perp}||^2 \sim \frac{4G^2(M+m)^2}{V^2b^2}-\frac{8G^2(M+m)^2}{c^2V^2b^2}\times\\
\times\bigg[(4+2\eta)\frac{G(M+m)}{b} +3V^2-\frac{7}{2}\eta\,V^2\bigg],
\end{multline}
where we have assumed that $\hat{\textbf{r}}\cdot \hat{\textbf{V}}\sim 1$ and dropped the terms proportional to $1/c^4$. The parallel velocity change of the test particle becomes
\begin{multline}
{\Delta \textbf{v}_{\rm T \parallel}} \sim -\frac{\mu}{M}\frac{||\Delta \mathbf{V}_{\perp}||^2}{2V^2} \textbf{V}\sim\\ \sim\bigg[-\frac{2mG^2(M+m)}{b^2V^4}+\frac{4G^3m(M+m)^2 (4+2\eta)}{c^2b^3V^4}+\\
+\frac{4G^2m(M+m)(3-\frac{7}{2}\eta)}{c^2b^2V^2} \bigg]\textbf{V},
\end{multline}
%%%%%%%%%%%%%%%%%%%%%%%%%%%%%%%%%%%%%%%%%%%%%%%%%%%%%%%%%%%%%%%%%%%%%%%%%%%%%%%%%%%%%%%
\begin{figure}
\includegraphics[width = \columnwidth]{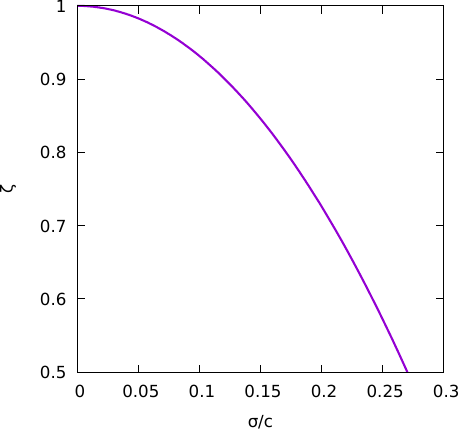}
\caption{Correcting factor $\zeta$ in the low-velocity limit as function of $\sigma$.}
\label{fig1pnclass}
\end{figure}
%%%%%%%%%%%%%%%%%%%%%%%%%%%%%%%%%%%%%%%%%%%%%%%%%%%%%%%%%%%%%%%%%%%%%%%%%%%%%%%%%%%%%%
so that, assuming again as in Section 2 that the number of encounters is given by (\ref{nenc}), the finite differences velocity change of particle $M$ is expressed as
\begin{multline}
\frac{\Delta \textbf{v}_{\rm T}}{\Delta t}= 2\pi bdbVnf(\textbf{v}_{\rm F})d^3 \textbf{v}_{\rm F}\bigg[-\frac{2mG^2(M+m)}{b^2V^4}\textbf{V}+\\
+\frac{4G^3m(M+m)^2 (4+2\eta)}{c^2b^3V^4}\textbf{V}+\frac{4G^2m(M+m)(3-\frac{7}{2}\eta)}{c^2b^2V^2} \textbf{V}\bigg].
\end{multline}
With the standard integration over the impact parameter $b$, we easily obtain 
\begin{multline}\label{1PNnaive}
\frac{d\textbf{v}_{\rm T }}{dt}=-4\pi m n G^2(M+m) \log \Lambda \int{  \frac{f(\textbf{v}_{\rm F})\textbf{V}}{V^3}}d^3\textbf{v}_{\rm F}+\\
+\frac{16\pi m n G^3(M+m)^2(2+\eta)}{\tilde{b}_{\rm min}c^2} \int{ \frac{f(\textbf{v}_{\rm F})\textbf{V}}{V^3}}d^3\textbf{v}_{\rm F}+\\
+\frac{8\pi m n G^2(M+m)(3-\frac{7}{2}\eta)\log\Lambda}{c^2}\int{ \frac{f(\textbf{v}_{\rm F})\textbf{V}}{V}}d^3\textbf{v}_{\rm F},
\end{multline}
{where the term $(b_{\rm max}-b_{\rm min})/b_{\rm min}b_{\rm max}$, arising from the integration in $b$, has been substituted in the second addendum with $\tilde{b}_{\rm min}^{-1}$. The latter is a {\it bona fide} minimum impact parameter defined by $\tilde{b}_{\rm min}=2G(M+m)/\sigma^2$, where $\sigma$ is the velocity dispersion of $f$.\\
\indent In practice, a naive $1$PN extension of the classical case, independently on the specific choice of $f(\mathbf{v}_{\rm F})$ augments the Chandrasekhar expression of the DF (first line of Eq. \ref{1PNnaive}) of two additional terms. The first one has a different dependence on the impact parameter, but the same integral on $\textbf{V}$, while the second contains the classical Coulomb logarithm, but a different integral on $\textbf{V}$. For the special case of a isotropic Maxwellian distribution with velocity dispersion $\sigma$, the 1PN DF expression can be easily integrated in the limit of $v_{\rm T}<\sigma$ and rewritten as
\begin{equation}\label{1PNnaive2}
\frac{d\textbf{v}_{\rm T }}{dt}=-4\pi \rho G^2(M+m) \frac{\log \Lambda}{\sigma^3}\zeta\textbf{v}_{\rm T },
\end{equation}
where the corrective factor $\zeta$, plotted in Fig. \ref{fig1pnclass} as function of $\sigma/c$ is defined by
\begin{equation}
\zeta=1-\left[\frac{8+4\eta}{\log \Lambda}+6-14\eta\right]\frac{\sigma^2}{c^2}.
\end{equation}
Interestingly, the net effect of a simple 1PN correction is to reduce the DF drag force with respect to its classical counterpart given by Eq. (\ref{dfchandra}). This has the relevant consequence that a highly massive object, for which the GR corrections to its gravitational field can not be neglected, (e.g. a SMBH moving at non-relativistic speed in a star system) suffers a less effective gravitational drag than what one would estimate using Chandrasekhar formula. This implies in that specific case an even longer in-spiral time-scale for the BHs in galactic cores. The reason for a lower DF effect in a simple 1PN approximation is ascribed to the fact that the relativistic precession due to the velocity dependent force term in a 2-body scatter, always acts on the opposite direction with respect to the deflection caused by the $1/r^2$ force term.\\
\indent The behaviour of the 1PN DF can not be extrapolated for large (relativistic) velocities, since for such large values of $v$ the velocity composition should be made in terms of Lorentz transforms as in Sect. 4.1.}
\subsection{Relativistic velocities}
We now extend the DF formula to the case where the particles velocities are relativistic and the density is such that the force is to be evaluated at 1PN order during close encounters.\\
\indent Following \cite{Lee:69}, we consider a mass $M$ moving at $\mathbf{v}_{\rm T}$ in a uniform medium of field stars, of constant number density $n$, with isotropic velocity distribution $f(\mathbf{v}_{\rm F})$, in an inertial frame $\textit{S}$. As usual, in a time interval $\Delta t$, its velocity will change by an amount $\sum\Delta\mathbf{v}_{\rm T}$ after $n_{\rm enc}$ encounters.\\
\indent As the effect of General Relativity will be accounted only during the deflection of the test particle $M$, its velocity will be transformed according to the Lorentz transformations of Special Relativity.\\
\indent As in the classical case, the isotropy of $f(\mathbf{v}_{\rm F})$ allows us to write
\begin{equation}
\sum\Delta \mathbf{v}_{\rm T} \wedge \mathbf{v}_{\rm T}= 0,
\end{equation}
so it is sufficient to perform
\begin{equation}
\sum\Delta{v}_{\rm T\parallel} = \frac{ \sum\Delta \mathbf{v}_{\rm T} \cdot \mathbf{v}_{\rm T} }{v_{\rm T}}.
\end{equation}
Let us consider a single encounter of a test star with a field star: the velocity $\mathbf{v}_{\rm T}$ will become $\mathbf{v}_{\rm T}^{ f}$, so we have
\begin{equation} \label{100}
\Delta {v}_{\rm T\parallel}= \frac{  (\mathbf{v}_{\rm T}^{ f}-\mathbf{v}_{\rm T}) \cdot \mathbf{v}_{\rm T} }{v_{\rm T}}\,\,.
\end{equation}
It is now useful to introduce a second reference frame, $\textit{S}^{\prime}$, in which the test star is, initially, at rest. Let $\textbf{w}_{\rm T}^{ f}$ be the velocity of the test star after the encounter, in the frame $\textit{S}^{\prime}$.
With our assumptions (the motion of all the stars will be approximately described by straight line) we can write, without loosing generality, that
\begin{equation}
\mathbf{v}_{\rm T}^{ f}= \frac{ (\textbf{w}_{\rm T}^{ f}+\mathbf{v}_{\rm T})}{ 1+ \frac{\textbf{w}_{\rm T}^{ f}\cdot \mathbf{v}_{\rm T}}{c^2}}\,\,,
\end{equation}
and therefore
\begin{equation}
\mathbf{v}_{\rm T}^{ f}\cdot \mathbf{v}_{\rm T}= \frac{ (\textbf{w}_{\rm T}^{ f}+\mathbf{v}_{\rm T})\cdot \mathbf{v}_{\rm T}}{ 1+ \frac{\textbf{w}_{\rm T}^{ f}\cdot \mathbf{v}_{\rm T}}{c^2}}\,\,.
\end{equation}
Noting that $\Delta \textbf{w}_{\rm T}=\textbf{w}_{\rm T}^{ f}$, (i.e., the test star is initially at rest in $\textit{S}^{\prime}$) and keeping only terms of order $1/c^2$, Eq. (\ref{100}) becomes after some algebra
\begin{multline}\label{101}
\begin{gathered}
\Delta {v}_{\rm T \parallel}= \frac{1}{v_{\rm T}} \Bigg[\frac{ (\textbf{w}_{\rm T}^{f}+\mathbf{v}_{\rm T})\cdot \mathbf{v}_{\rm T} } {1+ \frac{\textbf{w}_{\rm T}^{ f}\cdot \mathbf{v}_{\rm T}}{c^2} }  - v_{\rm T}^2  \Bigg]  \underset{\textbf{w}_{\rm T}^{f}=\Delta \textbf{w}_{\rm T}}{\approx}\\
\approx\frac{1}{v_{\rm T}}\Bigg[\Delta \textbf{w}_{\rm T} \cdot\mathbf{v}_{\rm T} \gamma_{v_{T}}^{-1}-\frac{(\Delta \textbf{w}_{\rm T} \cdot \mathbf{v}_{\rm T})^2}{c^2} \Bigg].
\end{gathered}
\end{multline}
We must now sum Eq.\,(\ref{101}) over all possible values of $\Delta\textbf{w}_{\rm T}$. We recall that  $\textbf{v}_{\rm F}$ and $\textbf{w}_{\rm F}$ are the initial velocities of field stars in the frames $\textit{S}$ and $\textit{S}^{\prime}$, related by the following Lorentz transformation (e.g. see \citealt{Landau})
\begin{equation}\label{vtrasc}
\mathbf{w}_{\rm F}=\frac{1}{1-\frac{\mathbf{v}_{\rm F} \cdot \mathbf{v}_{\rm \tau}}{c^2}}
\Bigg[\frac{\mathbf{v}_{\rm F}}{\gamma_{\rm v \tau}}-\mathbf{v}_{\rm \tau}+\frac{1}{c^2}\frac{\gamma_{\rm v \tau}(\mathbf{v}_{\rm F} \cdot \mathbf{v}_{\rm \tau})\mathbf{v}_{\rm \tau}}{\gamma_{\rm v \tau} +1}\Bigg],
\end{equation}
where $\mathbf{v}_{\rm \tau}$ is the translational velocity of $\textit{S}^{\prime}$ relative to $\textit{S}$, $\gamma_{\rm v \tau}=(1-\mathbf{v}_{\rm \tau}^{2}/c^2)^{-1/2}$ and $b$ is, as usual, the impact parameter of the encounter, defined in $\textit{S}^{\prime}$. We also have that $\textbf{w}_{\rm F} \cdot \textbf{b}=w_{\rm F}\,b\cos\varphi$ while $\Delta \textbf{w}_{\rm T}$, decomposed into its components parallel and perpendicular to $\textbf{w}_{\rm F}$, becomes
\begin{equation}\label{deltavt}
\Delta \textbf{w}_{\rm T}=\Delta w_{\rm T \parallel} \frac{\textbf{w}_{\rm F} }{w_{\rm F}}+ \Delta w_{\rm T \perp} \frac{\textbf{b}}{b}.
\end{equation}
The velocity distribution of field particles in $\textit{S}^{\prime}$ is (e.g. see \citealt{Landau}) 
\begin{equation}\label{distprime}
f^{\prime}(\textbf{w}_{\rm F})=\gamma_{v_{\rm T}} f(\mathbf{v}_{\rm F}) \Bigg(\frac{\gamma_{w_{\rm F}}}{\gamma_{v_{\rm F}}}\Bigg)^4\,,
\end{equation}
where $\mathbf{v}_{\rm F}=\mathbf{v}_{\rm F}(\textbf{w}_{\rm F}, \mathbf{v}_{\rm T})$. Let us denote with
\begin{equation}
\Delta t^{\prime}=\gamma_{v_{T}}^{-1} \Delta t
\end{equation}
the transformed time interval during which the encounters with the field particles are summed.\\
\indent It is important to note that $f^{\prime}(\textbf{w}_{\rm F})$ in the frame $\textit{S}^{\prime}$ is only homogeneous, but no more isotropic. As a consequence of this, the impact parameter $b$ and the deflection angle $\phi$ are randomly distributed for a {\it given} value of $\textbf{w}_{\rm F}$. Integrating Eq. (\ref{101}) over all values of $\Delta \textbf{w}_{\rm T}$ we obtain the formal result
\begin{multline} \label{coefffin}
\frac{d{v}_{\rm T \parallel}}{dt} =\int{\,n f^{\prime}( \textbf{w}_{\rm F}) \gamma_{v_{T}}^{-1}w_{\rm F}}d^3\textbf{w}_{\rm F} \int_{b_{\rm min}}^{b_{\rm max}}{2\pi b db}\,\times\\
\times\int_{0}^{2\pi} {\frac{1}{2\pi} \frac{1}{v_{\rm T}}}\Bigg[ \Delta \textbf{w}_{\rm T} \cdot \mathbf{v}_{\rm T} \gamma_{v_{T}}^{-2}-\frac{(\Delta \textbf{w}_{\rm T} \cdot \mathbf{v}_{\rm T})^2}{c^2} \Bigg]d\varphi\,\,.
\end{multline}
The quantity $\Delta \textbf{w}_{\rm T}(\textbf{w}_{\rm F}, b)$ appearing in the equation above must be expressed as a function of $\textbf{w}_{\rm F}$ and $b$ and can be computed with the help of the two body problem 1PN-Lagrangian. For this purpose, it is convenient to introduce a third reference frame  $\textit{S}^{\prime\prime}$, in which the center of mass (c.o.m.) of the encounter is at rest at the origin of coordinates\footnote{As the location of the origin will, of course, depend on the masses and velocities of both stars in $S$, we must obviously consider a new frame for each encounter.}, where we denote the particles velocities by $\textbf{u}$.\\
\indent At 1PN order the equations of motion for the two-body encounter are usually derived from the Einstein-Infeld-Hoffmann Lagrangian (EIH, see \citealt{EIH}, see also \citealt{1938RSPSA.166..465E})
\begin{multline}\label{EIH}
\qquad\mathcal{L}_{\rm EIH}= \frac{1}{2} m \textbf{u}^{2}_{\rm F}+\frac{1}{2} M \textbf{u}^{2}_{\rm T} + \frac{1}{8c^2}(m\textbf{u}^{4}_{\rm F}+M\textbf{u}^{4}_{\rm T})+\frac{GmM}{r}\times\\
\times\Bigg\{1- \frac{1}{2c^2}\bigg[7(\textbf{u}_{\rm F} \cdot \textbf{u}_{\rm T})+(\textbf{u}_{\rm F} \cdot \textbf{n})(\textbf{u}_{\rm T} \cdot \textbf{n})
-3(\textbf{u}_{\rm F}-\textbf{u}_{\rm T})^{2} \bigg] -\frac{G(m+M)}{2rc^2} \Bigg\}\,\,.
\end{multline}
Remarkably, qualitatively similar (but slightly simpler) equations of motion can be derived using the gravitational analogous of the Darwin Lagrangian of electrodynamics (see e.g. \citealt{Jackson}, see also \citealt{2007EL.....7960002E,2014JGrav.2014E..1E}), often referred as Fock Lagrangian (see \citealt{1964tstg.book.....F,1972AmJPh..40...63K,deruelle})
\begin{multline}\label{Darwin}
\mathcal{L}_{\rm Darwin}= \frac{1}{2} m \textbf{u}^{2}_{\rm F}+\frac{1}{2} M \textbf{u}^{2}_{\rm T} + \frac{1}{8c^2}(m\textbf{u}^{4}_{\rm F}+M\textbf{u}^{4}_{\rm T})+ \frac{GmM}{r}\times \\
\times\Bigg[1- \frac{1}{2c^2}\bigg(\textbf{u}_{\rm F} \cdot \textbf{u}_{\rm T}+(\textbf{u}_{\rm F} \cdot \textbf{n})(\textbf{u}_{\rm T} \cdot \textbf{n}) \bigg)\Bigg]\,\,.
\end{multline}
In Eqs. (\ref{EIH}) and (\ref{Darwin}), $\textbf{r}=\textbf{r}_{\rm F}-\textbf{r}_{\rm T}$ is the (instantaneous) relative position vector and $\textbf{n}={\textbf{r}}/{||\textbf{r}||}$, (see \citealt{Fagu,DD,Zuc}, for details). In both cases, the c.o.m. coordinates at the 1PN order are
\begin{equation}
\textbf{r}_{\rm cm}=\frac{\mathcal{E}_{\rm T}\textbf{r}_{\rm T}+\mathcal{E}_{\rm F}\textbf{r}_{\rm F}}{\mathcal{E}_{\rm T}+\mathcal{E}_{\rm F}},
\end{equation}
where
\begin{eqnarray}
\mathcal{E}_{\rm T}=Mc^2+\frac{1}{2}M\textbf{u}^{2}_{\rm T}-\frac{1}{2}\frac{GMm}{||\textbf{r}_{\rm T}-\textbf{r}_{\rm F}||},\nonumber\\
\mathcal{E}_{\rm F}=mc^2+\frac{1}{2}m\textbf{u}^{2}_{\rm F}-\frac{1}{2}\frac{GMm}{||\textbf{r}_{\rm T}-\textbf{r}_{\rm F}||}.
\end{eqnarray}
Since the c.o.m is by definition at rest at the origin of $\textit{S}^{\prime\prime}$, (i.e. $\textbf{r}_{\rm cm}=0$), in terms of the relative velocity $\textbf{u}$ we can always write
\begin{equation}\label{uf1PN}
\textbf{u}_{\rm T} = \frac{m}{m+M}\textbf{u} + \mathcal{O}(1/c^2);\quad \textbf{u}_{\rm F} = -\frac{M}{m+M}\textbf{u} + \mathcal{O}(1/c^2),
\end{equation}
so that
\begin {equation}
\frac{1}{2}(Mu^{2}_{\rm T}+mu^{2}_{\rm F})=\frac{1}{2} \bigg( {\frac{mM}{M+m} }u^{2}  \bigg) +\mathcal{O}(1/c^4).
\end{equation}
At variance with \cite{Lee:69}, we will assume hereafter the Lagrangian (\ref{Darwin}), that in terms of the relative velocity $\mathbf{u}$, once conveniently transformed in polar coordinates, becomes
\begin{multline}
\mathcal{L}_{\rm D}=\frac{1}{2} \mu\bigg[\bigg(\frac{dr}{dt} \bigg)^2+r^2 \bigg(\frac{d\vartheta}{dt}\bigg)^2\bigg] +\, \Phi \,+ \frac{1}{8c^2} \mu\bigg(\frac{\mu}{\mu_{\rm 3}}\bigg)^3 \times\\
\times\,\bigg[\bigg(\frac{dr}{dt} \bigg)^2+r^2 \bigg(\frac{d\vartheta}{dt}\bigg)^2\bigg]^2+\frac{\Phi}{2c^2}\frac{\mu}{\mathcal{M}} \bigg[2\bigg(\frac{dr}{dt} \bigg)^2+r^2 \bigg(\frac{d\vartheta}{dt}\bigg)^2\bigg].
\end{multline}
In the equation above, $\mu$ is again the reduced mass, $\mathcal{M}=M+m$, $\mu_{\rm 3}=[m^3M^3/(m^3+M^3)]^{1/3}$, and $\Phi={G\mu\mathcal{M}}/{r}$ is the Newtonian gravitational potential energy.  With such choice, the associated particles equation of motion are obtained in explicit form at the order $1/c^2$, while conversely, in the EIH case the force of particle $M$ on particle $m$ depends on the force of particle $m$ on particle $M$.\\
\indent In analogy with the non relativistic velocity case discussed above, we now obtain (see Appendix \ref{app2} for the explicit mathematical details) the DF formula in integral form as 
\begin{multline}\label{pippo1}
\frac{d{v}_{\rm T \parallel}}{dt}=\int{\,n f^{\prime}( \textbf{w}_{\rm F})} \gamma_{v_{\rm T}}^{-1} w_{\rm F}d^3\textbf{w}_{\rm F}\int_{b_{\rm min}}^{b_{\rm max}}{2\pi b db}\\
\int_{0}^{2\pi}{\frac{1}{2\pi} }\frac{1}{v_{\rm T}}\Bigg\{ \gamma_{v_{\rm T}}^{-2}\bigg[ \frac{\mathbf{v}_{\rm T} \cdot \textbf{w}_{\rm F}}{w_{\rm F}}  \Delta w_{\rm T \parallel} + \frac{\mathbf{v}_{\rm T} \cdot \textbf{b}}{b} \Delta w_{\rm T \perp} \bigg]+\\
- \frac{1}{c^2} \bigg[  \bigg(\frac{\mathbf{v}_{\rm T} \cdot \textbf{w}_{\rm F}}{w_{\rm F}} \bigg)^2  \big( \Delta w_{\rm T \parallel} \big)^2-2\bigg(\frac{\mathbf{v}_{\rm T} \cdot \textbf{w}_{\rm F}}{w_{\rm F}} \bigg) \bigg(\frac{\mathbf{v}_{\rm T} \cdot \textbf{b}}{b} \bigg) \Delta w_{\rm T \parallel}  \Delta w_{\rm T \perp}+\\
+\bigg(\frac{\mathbf{v}_{\rm T} \cdot \textbf{b}}{b} \bigg)^2  \big(  \Delta w_{\rm T \perp} \big)^2  \bigg]\Bigg\}d\varphi.
\end{multline}
We note that, all the linear terms in the equation above containing ${\mathbf{v}_{\rm T} \cdot \textbf{b}}/{b}$ vanish when they are integrated in $d\varphi$. Strictly speaking, $\varphi$ would be the angle between $\mathbf{w}_{\rm F}$ and $\mathbf{b}$. As in \cite{Lee:69} we assume, however, that it is also the angle between $\mathbf{v}_{\rm T}$ and $\mathbf{b}$, while the quadratic terms yield 
\begin{multline}
\begin{gathered}
\int_{0}^{2\pi}{\frac{1}{2\pi}}  \bigg(\frac{\mathbf{v}_{\rm T} \cdot \textbf{b}}{b} \bigg)^2 d\varphi =  \bigg(\frac{{v}_{\rm T}{b}}{b} \bigg)^2 \int_{0}^{2\pi}{\frac{\cos^2 \varphi}{2\pi}}d\varphi=\frac{(v_{\rm T}b)^2}{2b^2}=\\
= \frac{1}{2}\bigg(\frac{v_{\rm T}w_{\rm F}}{w_{\rm F}}\bigg)^2=\frac{1}{2} \bigg( \frac{v_{\rm T}w_{\rm F}\sin\varphi}{w_{\rm F}} \bigg)^2 = \frac{1}{2} \bigg( \frac{ \mathbf{v}_{\rm T} \wedge \textbf{w}_{\rm F} }{w_{\rm F}}   \bigg)^2 .
\end{gathered}
\end{multline}
In the equation above, we have used the fact that $\textbf{w}_{\rm F}\cdot \textbf{b}=0$, (i.e. $\textbf{w}_{\rm F}$ is perpendicular to $\textbf{b}$ and then $\sin\varphi=1$). We then rewrite Eq. (\ref{pippo1}) as
\begin{multline} \label{DFpost}
\begin{gathered}
\frac{d{v}_{\rm T \parallel}}{dt}=\int{\,n f^{\prime}(\textbf{w}_{\rm F})}\gamma_{v_{\rm T}}^{-1} \frac{w_{\rm F}}{v_{\rm T}}d^3\textbf{w}_{\rm F} \int_{b_{\rm min}}^{b_{\rm max}}{2\pi b}\,\,\, \times \\
\times \Bigg\{ \gamma_{v_{\rm T}}^{-2} \bigg(\frac{\mathbf{v}_{\rm T} \cdot \textbf{w}_{\rm F}}{w_{\rm F}}\bigg) \Delta w_{\rm T \parallel} -\frac{1}{c^2} \bigg[\bigg(\frac{\mathbf{v}_{\rm T} \cdot \textbf{w}_{\rm F}}{w_{\rm F}}\bigg)^2 (\Delta w_{\rm T \parallel})^2 +\\
+\frac{1}{2} \bigg( \frac{ \mathbf{v}_{\rm T} \wedge \textbf{w}_{\rm F} }{w_{\rm F}}   \bigg)^2  (\Delta w_{\rm T \perp})^2 \bigg] \Bigg\}db.
\end{gathered}
\end{multline}
In order to perform the integration over the impact parameter, otherwise hardly feasible, we may neglect all the terms that decrease faster than ${1}/{b}$, as $b\rightarrow +\infty$ using the so called dominant approximation. By doing so, dropping all terms proportional to $(\Delta w_{\rm T \parallel})^2$ we obtain
\begin{multline}
\frac{d{v}_{\rm T \parallel}}{dt}= \int{\, n f^{\prime}( \textbf{w}_{\rm F})}\gamma_{v_{\rm T}}^{-1} \frac{w_{\rm F}}{v_{\rm T}} d^3\textbf{w}_{\rm F}
\int_{b_{\rm min}}^{b_{\rm max}}{2\pi b}\times\\
\times \Bigg\{ \gamma_{v_{\rm T}}^{-2} \bigg(\frac{\mathbf{v}_{\rm T} \cdot \textbf{w}_{\rm F}}{w_{\rm F}}\bigg)\Delta w_{\rm T \parallel} -\frac{1}{2} \bigg( \frac{ \mathbf{v}_{\rm T} \wedge \textbf{w}_{\rm F} }{w_{\rm F}}   \bigg)^2  (\Delta w_{\rm T \perp})^2 \Bigg\}db.
\end{multline}
This established, we now have to evaluate $\Delta w_{\rm T \parallel}$ and $(\Delta w_{\rm T \perp})^2$ (the finite parallel and perpendicular velocity changes) as function (see Appendix \ref{app3}) of the deflection angles. We obtain, at 1PN the following expressions 
\begin{gather}\label{pippo77}
\Delta w_{\rm T \parallel}\approx -2\textit{u}_{\rm T} \bigg\{\frac{1}{1+b^2/\mathcal{R}^2}+ \frac{\textit{u}^2}{c^2}\frac{1}{1+b^2/\mathcal{R}^2}+\nonumber\\
\qquad\qquad\qquad\qquad\qquad+\frac{\textit{u}^2 b^2}{c^2 \mathcal{R}^2}\Big(\frac{\mu}{\mathcal{M}}-\frac{\mu^3}{\mu_{\rm 3}^3}\Big) \frac{1}{(1+b^2/\mathcal{R}^2)^2}\bigg\} \,,\\
(\Delta w_{\rm T \perp})^2\approx \textit{u}^{2}_{\rm T} \frac{4b^2}{\mathcal{R}(1+b^2/\mathcal{R}^2)^2}.
\end{gather}
Substituting Eqs.(\ref{pippo77}) in Eq. (\ref{DFpost}) and performing the integral over $b$ keeping only the terms yielding the Coulomb logarithm in
\begin{multline}
\begin{gathered}
\int_{b_{\rm min}}^{b_{\rm max}}{2\pi b} \Delta w_{\rm T \parallel}db = -2\pi \textit{u}_{\rm T}\int_{b_{\rm min}}^{b_{\rm max}}{2b}\Bigg[\frac{1}{1+b^2/\mathcal{R}^2}+ \frac{\textit{u}^2}{c^2}\times\\
\,\,\quad\times\frac{1}{1+b^2/\mathcal{R}^2}
+\frac{\textit{u}^2 b^2}{c^2 \mathcal{R}^2}\Big(\frac{\mu}{\mathcal{M}}-\frac{\mu^3}{\mu_{\rm 3}^3}\Big) \frac{1}{(1+b^2/\mathcal{R}^2)^2}\Bigg]db\approx\\ \approx -2\pi \textit{u}_{\rm T} \mathcal{R}^2
 \log{\bigg(\frac{1+b^{2}_{\rm max}/\mathcal{R}^2}{1+b^{2}_{\rm min}/\mathcal{R}^2}\bigg)}\bigg[1+\frac{\textit{u}^2}{c^2}+\frac{\textit{u}^2}{c^2}\bigg(\frac{\mu}{\mathcal{M}}-\frac{\mu^3}{\mu_{\rm 3}^3}\bigg)\bigg]\,\,,
\end{gathered}
\end{multline}
gives
\begin{multline}
\begin{gathered}
\int_{b_{\rm min}}^{b_{\rm max}}{2\pi b} (\Delta w_{\rm T \perp})^2 db \approx 4 \textit{u}^{2}_{\rm T}\, \pi \int_{b_{\rm min}}^{b_{\rm max}} { \frac{2b^3 \, db}{\mathcal{R}^2 (1+b^{2} / \mathcal{R})^2 }} \approx\\
\approx 4 \textit{u}^{2}_{\rm T} \pi \mathcal{R}^2  \log{\bigg(\frac{1+b^{2}_{\rm max}/\mathcal{R}^2}{1+b^{2}_{\rm min}/\mathcal{R}^2}\bigg)}.
\end{gathered}
\end{multline}
{After some further algebraic manipulation, we finally obtain the 1PN dynamical friction formula in compact form as
\begin{multline}\label{ddff}
\begin{gathered}
\frac{d{v}_{\rm T \parallel}}{dt}=-4\pi G^2(M+m)\rho\frac{\log\Lambda}{\gamma_{v_{\rm T}}v_{\rm T}}\int  \frac{f^{\prime}(\textbf{w}_{\rm F})}{\bigg[1+\frac{4mMw_{\rm F}^2}{(M+m)^2c^2}\bigg]w_{\rm F}^3}\\
\Bigg\{ \frac{\mathbf{v}_{\rm T} \cdot \textbf{w}_{\rm F}}{\gamma_{v_{\rm T}}^{2}}\bigg[1+\frac{(9Mm+m^2)w_{\rm F}^2}{2(M+m)^2c^2}\bigg] +  \frac{m^2 (\mathbf{v}_{\rm T} \wedge \textbf{w}_{\rm F})^2}{(M+m)^2c^2}   \Bigg\}d^3\textbf{w}_{\rm F},
\end{gathered}
\end{multline}
where again we have assumed the velocity averaged Coulomb logarithm.\\ 
\indent Equation (\ref{ddff}), although highly simplified by all the assumptions made above, represents the most general relativistic derivation of the dynamical friction formula. In analogy with the treatment of the special relativistic DF formula, we have investigated numerically the behaviour of Eq. (\ref{ddff}) against its classical counterparts in some cases of astrophysical interest by solving the integral appearing in the equation above for the relativistic $f(v)$ given again by Eq. (\ref{juttner}).\\
\indent In Figure \ref{fig1pnmbig} we show the DF force acting on the test particle $M$ as function of its velocity for a  $10^{5}M_{\odot}$ and a $10^{7}M_{\odot}$ black holes moving respectively in a dense star cluster and a galactic core. In both cases, for $v_{\rm T}<\sigma$ the classical expression slightly overestimates the drag force acting on $M$, thus confirming the small $v$ estimates of Sect. 5.1 (crf. Fig \ref{fig1pnclass}). For large velocities the extra terms containing dot and cross products of the test particle velocity and the relative velocity $w_{\rm F}$ in $S^{\prime\prime}$ become dominant, thus yielding a much higher dynamical friction with respect to the Chandrasekhar formula.\\
%%%%%%%%%%%%%%%%%%%%%%%%%%%%%%%%%%%%%%%%%%%%%%%%%%%%%%%%%%%%%%%%%%%%%%%%%%%%%%%%%%%%%%%
\begin{figure*}
\includegraphics[width = 0.8\textwidth]{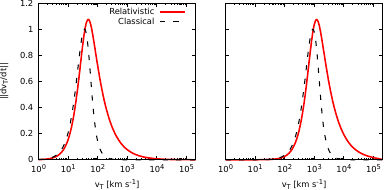}
\caption{Relativistic and classical dynamical friction as function of $v_{\rm T}$ evaluated for a $10^{5}M_{\odot}$ IMBH in a star cluster with $\sigma=25$ km s$^{-1}$ and a $10^{7}M_{\odot}$ SMBH in a galactic core with $\sigma=500$ km s$^{-1}$}
\label{fig1pnmbig}
\end{figure*}
%%%%%%%%%%%%%%%%%%%%%%%%%%%%%%%%%%%%%%%%%%%%%%%%%%%%%%%%%%%%%%%%%%%%%%%%%%%%%%%%%%%%%%
\indent Finally, for reasons of completeness it is worth to discuss its behaviour in its two fundamental limits; i.e. $c\rightarrow\infty$ and $M\gg m$. In the limit $c\rightarrow \infty$, Eq. (\ref{ddff}) becomes
\begin{equation}
\frac{d{v}_{\rm T \parallel}}{dt}=-4\pi G^2(M+m)\rho\ \frac{\log\Lambda}{v_{\rm T}}\int  \frac{f(\textbf{v}_{\rm F})}{{w}_{\rm F}^3}(\mathbf{v}_{\rm T} \cdot \textbf{w}_{\rm F}) d^3\textbf{w}_{\rm F},   
\end{equation}
where we have used Eq. (\ref{distprime}) to re-obtain $f(v_{\rm F})$ from $f(w_{\rm F})$. In order to retrieve Eq.(\ref{13}), one has to substitute $\mathbf{w}_{\rm F}$ with $\mathbf{v}_{\rm F}$, by using Eq.(\ref{vtrasc}) in its classical limit (i.e. the classical velocity summation rule $\mathbf{w}_{\rm F}=\mathbf{v}_{\rm F}-\mathbf{v}_{\rm \tau}$) choosing $\mathbf{v}_{\rm \tau}=\mathbf{v}_{\rm T}$. Once making explicit $\mathbf{v}_{\rm T} \cdot \textbf{w}_{\rm F}$ as $v_{\rm T}w_{\rm F}\cos\phi$, the test particle velocity $v_{\rm T}$ cancels out with the same factor outside the integral and then, in the assumption of isotropy for $f_{v_{\rm F}}$ the same considerations about the angle $\phi$ between $\mathbf{v}_{\rm F}$ and $\mathbf{v}_{\rm T}$ used in Sect. 4.1 can be made to retrieve Eq. (\ref{dfchandra}) in vectorial form.\\
\indent In the limit of very large test particle mass $M$ we obtain 
\begin{multline}\label{final}
\begin{gathered}
\frac{d{v}_{\rm T \parallel}}{dt}=-4\pi G^2M\rho\frac{\log\Lambda}{v_{\rm T}\gamma_{v_{\rm T}}^{3}}\int\frac{f^{\prime}(\mathbf{w}_{\rm F})(\mathbf{v}_{\rm T} \cdot \mathbf{w}_{\rm F})}{w_{\rm F}^3}\frac{1+9\xi/2}{1+4\xi}d^3\textbf{w}_{\rm F},
\end{gathered}
\end{multline}
where $\xi=mw_{\rm F}^2/Mc^2$ and all terms proportional to $(m/M)^2$ have been discarded. Curiously, the multiplicative factor containing $\xi$ in Equation (\ref{final}) is roughly unity when $m$ is vanishingly small, which implies that, at least at low (relative) velocities $w_{\rm F}$ the mass dependent terms in Eq. (\ref{ddff}) can be neglected, while in other cases it still bares an explicit dependence on $M$ and $m$ inside the integral, differently from what one would obtain in the same limit for the classical expression, (see the discussion in \citealt{BinTre:08} and \citealt{Ciotti:20}). In other words, in a relativistic set-up the dynamical friction acting on a massive particle $M$ is slightly different in two systems with the {\it same} mass density $\rho$ but different field particle mass $m$. However, for large $M/m$, as in the case for example of a star moving in a gas of relativistic dark matter particles with typical masses in the range of those of elementary particles, the specific value of $m$ is immaterial once $\rho$ is known.\\
\indent Moreover, we notice that Eq. (\ref{final}) has a $1/\gamma_{v_{\rm T}}^{3}$ dependence on the Lorentz factor, instead of $1/\gamma_{v_{\rm T}}$ as in the weak scattering limit for a large test mass in a background of particles of infinitesimally small masses $m$, (cf. Eq. 2.26 of \citealt{Syer:94}).}
%%%%%%%%%%%%%%%%%%%%%%%%%%%%%%%%%%%%%%%%%%%%%%%%%%%%%%%%%%%%%%%%%%%%%%%%%%%
\section{Summary and perspectives}
%%%%%%%%%%%%%%%%%%%%%%%%%%%%%%%%%%%%%%%%%%%%%%%%%%%%%%%%%%%%%%%%%%%%%%%%%%
In this preparatory work on the the dynamical friction in relativistic systems, we have explored two relevant cases, the one involving relativistic velocity distribution and classical forces between particles, and the other with strong scattering with and without relativistic velocities.\\
\indent We find that, extending the asymptotic expression of \cite{Syer:94} valid only valid for scattering angles to a generic $\theta_{\rm defl}$ and for the case of a pure Newtonian $1/r^2$ force law, a particle moving through a medium with a relativistic velocity distribution suffers a slightly larger drag with respect to the what is evaluated with the standard classical Chandrasekhar approach, since all particle contribute in the slowing down of the test mass (not only those with $v_{\rm F}<v_{\rm T}$).\\
\indent Remarkably, we also found out that a naive generalization of the classical DF formula in the case of strong scattering in post Newtonian approximation and non relativistic velocity distribution gives a smaller drag on the test particle with respect to the original Chandrasekhar expression due to the competing effects of orbit deflection and relativistic precession. A more complete treatment of relativistic DF using the gravitational Darwin Lagrangian (in lieu of the more complex Einstein-Infeld-Hoffmann Lagrangian) appears to confirm this fact. However, for $v_{\rm T}$ larger than the peak value of the velocity distribution, the behaviour is reversed and the relativistic DF expression dominates over its parent classical formulation. This  result appears to be confirmed by preliminary direct $N-$body simulations (to be published elsewhere, see \citealt{chiari2}), where we have studied the orbital decay of a $10^5M_{\odot}$ BH placed initially on a circular orbit in a star cluster with and without the post Newtonian correction to the force law, finding a slightly (of a factor 1.1) larger in-spiraling time in the runs with the post Newtonian corrections.\\
\indent So far, the results discussed in the present paper could be relevant for the dynamics of massive objects at the centre of dense star clusters where both strong deflections and large velocity dispersion  may occur, as well as models of hot Dark Matter with relativistic velocities. It is worth to mention that the formalism used in Sect. 4.2 could be also extended to treat relativistic charged particles (in the limit of negligible radiation losses) using the electrodynamical Darwin Lagrangian. In this respect, it should also be stressed again that the expression derived in Sect. 4.1 can also be used in plasma physics, as the velocity and impact parameter integrals are exactly the same as in the DF friction formula in charged particle systems. In particular, as already mentioned above, in such environments fat-tailed power-law velocity distributions are rather common. Therefore, due to the large contribution of such tails in the relativistic velocity integrals we may expect relevant discrepancies with the classical predictions.\\
\indent At this point, it remains to be determined whether the inclusion of explicitly dissipative terms (i.e. the effect of momentum loss due to the emission of gravitational waves) would alter significantly the relativistic dynamical friction drag. To do so, in principle one should include in Eq. (\ref{postn}) all terms up to the order 2.5. Moreover, an additional generalization that seems to be feasible could be in the direction of systems with a mass spectrum with mass-dependent average relativistic factor, in other words, where for example only the low mass particles have relativistic velocities.\\
\indent In this paper we have explored only infinitely extended systems in the spirit of the original treatment of the dynamical friction. As mentioned above star cluster simulations are in the works. In the next paper of this series, we will investigate by means of post Newtonian $N-$body simulations the collisional dynamics of compact objects kicked by gravitational waves emission in dense stellar systems and the relativistic corrections on their dynamical friction induced retention. 
\begin{acknowledgements}
We thank Lapo Casetti, for the useful discussions at an early stage of this work. One of us (PFDC) wishes to acknowledge partial financial support from the MIUR-PRIN2017 project \textit{Coarse-grained description for non-equilibrium systems and transport phenomena (CO-NEST)} n.\ 201798CZL. 
\end{acknowledgements}
%%%%%%%%%%%%%%%%%%%%%%%%%%%%%%%%%%%%%%%%%%
   \bibliographystyle{aa} % style aa.bst
   \bibliography{biblio} % your references Yourfile.bib
\begin{appendix} %First appendix
\section{Post-Newtonian dynamical friction and the (gravitational) Darwin Lagrangian}\label{app2}
When computing the deflection for an isolated encounter, we need to express, as done classically the equation of motion for the reduced particle. Instead of deriving such equations by using the principle of least action, it is easier to find their first integrals as
\begin{equation}
p_{\rm \vartheta}  = \frac{\partial \mathcal{L}_{\rm D}}{\partial(\partial\vartheta/\partial t)};\quad 
E  =p_{\rm \vartheta} \frac{d\theta}{dt}+p_{\rm r} \frac{dr}{dt}-\mathcal{L}_{\rm D}\equiv \mathcal{H}_{\rm D}, 
\end{equation}
where $\mathcal{H}_{\rm D}$ is the Darwin Hamiltonian, in analogy with the electromagnetic case. At the lowest order, (i.e. when $\mathcal{L}_{\rm Darwin}\equiv \mathcal{L}_{\rm Newton}$), we have
\begin{equation}
p_{\rm \vartheta}  = \mu r^2\frac{d\vartheta}{dt},\quad \mathcal{E}  = \frac{1}{2} \mu\bigg[\bigg(\frac{dr}{dt} \bigg)^2+r^2 \bigg(\frac{d\vartheta}{dt}\bigg)^2\bigg] -\, \Phi \,.
\end{equation}
These expressions will be used to eliminate $dr/dt$ and $d\vartheta/dt$ from $p_{\rm \vartheta}$ and $\mathcal{E}$ in all terms of order $1/c^2$. This is possible because $\mathcal{L}_{\rm Darwin}=\mathcal{L}_{\rm Newton}+ {1}/{c^2}(...)$ and all terms in parentheses, already containing a factor of order $1/c^2$, are multiplied by another $1/c^2$ factor, out of parentheses, therefore adding up to the terms in $1/c^4$, that neglected in the 1PN approximation, (a more sophisticated proof of this argument is given, for example, in \citealt{DD}).\\
\indent At 1PN we then find
\begin{multline} \label{ptheta}
p_{\rm \vartheta} = \mu r^2 \frac{d\vartheta}{dt}\bigg\{1+\frac{1}{\mu c^2}\bigg[\mathcal{E}\frac{\mu^3}{\mu_{\rm 3}^3} +\Phi \bigg(\frac{\mu^3}{\mu_{\rm 3}^3} +\frac{\mu}{\mathcal{M}} \bigg)  \bigg] \bigg\},
\end{multline}
from which 
\begin{multline}\label{dtheta}
\frac{d\vartheta}{dt} =\frac{p_{\rm \vartheta}}{\mu r^2} \frac{1}{\{1+......\}}  \underset{\rm \sim\frac{1}{c^2}} {\approx}  \frac{p_{\rm \vartheta}}{\mu r^2} \bigg\{1- \frac{1}{\mu c^2}\bigg[\mathcal{E}\frac{\mu^3}{\mu_{\rm 3}^3}+\Phi \bigg(\frac{\mu^3}{\mu_{\rm 3}^3} +\frac{\mu}{\mathcal{M}} \bigg)\bigg]\bigg\},
\end{multline}
and 
\begin{multline} 
p_{\rm r} = \mu \frac{dr}{dt} \Bigg\{1+\frac{1}{\mu c^2} \Bigg[\mathcal{E}\frac{\mu^3}{\mu_{\rm 3}^3} +\Phi \bigg(\frac{\mu^3}{\mu_{\rm 3}^3} +\frac{\mu}{\mathcal{M}} \bigg)  \Bigg] \Bigg\}.
\end{multline}
In order to find $\mathcal{H}_{\rm D}$ as a function only of $r$ and $dr/d\vartheta$, we have to replace
\begin{equation}
\frac{dr}{dt}=\frac{dr}{d\vartheta}\frac{d\vartheta}{dt}
\end{equation}
in $p_{\rm r}$ and $\mathcal{L}_{\rm D}$, while $d\vartheta/dt$ is given by (\ref{dtheta}).
Therefore, keeping only terms of order $1/c^2$, we obtain
\begin{multline} 
\mathcal{L}_{\rm D}= \frac{1}{2}\frac{p^{2}_{\rm \vartheta}}{\mu r^4}\bigg(\frac{dr}{d\vartheta}\bigg)^2+\frac{1}{2}\frac{p^{2}_{\rm \vartheta}}{\mu r^2}+ \Phi + \frac{\Phi}{\mathcal{M}c^2}\bigg(\frac{dr}{d\vartheta}\bigg)^2 \frac{p^{2}_{\rm \vartheta}}{\mu r^4} 
-\frac{1}{\mu c^2}\times\\
\times\,\Bigg\{\frac{3}{2}\mathcal{E}^2 \frac{\mu^3}{\mu_{\rm 3}^3} +3 \mathcal{E}\Phi \frac{\mu^3}{\mu_{\rm 3}^3} +\mathcal{E}\Phi  \frac{\mu}{\mathcal{M}}+ \frac{3}{2} \Phi^2   \frac{\mu^3}{\mu_{\rm 3}^3}    +\Phi^2 \frac{\mu}{\mathcal{M}} \Bigg\},
\end{multline}
and finally
\begin{multline} \label{Hdarwin}
\begin{gathered}
\mathcal{H}_{\rm D}=p_{\rm \vartheta} \frac{d\vartheta}{dt}+p_{\rm r} \frac{dr}{dt}-\mathcal{L}_{\rm D}= \frac{1}{2} \frac{p^{2}_{\rm \vartheta}}{\mu r^4} \bigg(\frac{dr}{d\vartheta}\bigg)^2 \bigg(1-\frac{2\Phi}{\mathcal{M}c^2} \bigg)+\\
+\frac{1}{2} \frac{p^{2}_{\rm \vartheta}}{\mu r^2} \bigg(1-\frac{2\Phi}{\mathcal{M}c^2} \bigg)
- \Phi - \frac{1}{\mu c^2} \Bigg[\frac{1}{2} \mathcal{E}^2 \frac{\mu^3}{\mu_{\rm 3}^3} +  \mathcal{E} \Phi\,\,\times\\
\times \bigg( \frac{\mu^3}{\mu_{\rm 3}^3} - \frac{\mu}{\mathcal{M}}\bigg) + \frac{1}{2} \Phi^2  \bigg( \frac{\mu^3}{\mu_{\rm 3}^3} - 2\frac{\mu}{\mathcal{M}}\bigg)  \Bigg]= E.
\end{gathered}
\end{multline}
At this point, we need to express the azimuthal angle, $\vartheta$, in $\textit{S}^{\prime\prime}$. In order to do so, let us manipulate the Equation (\ref{Hdarwin}), by isolating the terms in $d\vartheta$ and $dr/r$ obtaining
\begin{multline} 
\Bigg[\frac{1}{2} \frac{p^{2}_{\rm \vartheta}}{\mu r^4}\bigg(\frac{dr}{d\vartheta}\bigg)^2 +\frac{1}{2} \frac{p^{2}_{\rm \vartheta}}{\mu r^2}  \Bigg] \bigg(1-\frac{2\Phi}{\mathcal{M}c^2} \bigg)= E+\Phi+\frac{1}{\mu c^2} \Bigg[\frac{1}{2} \mathcal{E}^2 \frac{\mu^3}{\mu_{\rm 3}^3}+\\
\mathcal{E} \Phi \bigg(\frac{\mu^3}{\mu_{\rm 3}^3} - \frac{\mu}{\mathcal{M}}\bigg) + \frac{1}{2} \Phi^2  \bigg( \frac{\mu^3}{\mu_{\rm 3}^3} - 2\frac{\mu}{\mathcal{M}}\bigg)\Bigg],
\end{multline}
from which, with a little rearrangement of terms we then get
\begin{multline} \label{rc}
\frac{1}{2} \frac{p^{2}_{\rm \vartheta}}{\mu r^4}\bigg(\frac{dr}{d\vartheta}\bigg)^2=\Bigg\{\frac{E+\Phi}{\big(1-\frac{2\Phi}{\mathcal{M}c^2} \big)} - \frac{1}{2} \frac{p^{2}_{\rm \vartheta}}{\mu r^2} 
+ \frac{1}{\mu c^2} \frac{1}{\big(1-\frac{2\Phi}{\mathcal{M}c^2} \big)}\times\\
\times\Bigg[\frac{1}{2} \mathcal{E}^2 \frac{\mu^3}{\mu_{\rm 3}^3} + \mathcal{E}\, \Phi \bigg( \frac{\mu^3}{\mu_{\rm 3}^3} - \frac{\mu}{\mathcal{M}}\bigg) + \frac{1}{2} \Phi^2  \bigg( \frac{\mu^3}{\mu_{\rm 3}^3} - 2\frac{\mu}{\mathcal{M}}\bigg)  \Bigg] \Bigg\}.
\end{multline}
The angle $\vartheta$ is obtained in integral form as
\begin{multline}\label{net}
\begin{gathered}
\vartheta=2\int_{r_{\rm c}}^{r} {\frac{1}{r} } \Bigg\{ \frac{2\mu r^2}{p_{\rm \vartheta}}\Bigg[ \frac{E+\Phi}{1-\frac{2\Phi}{\mathcal{M}c^2}} - \frac{1}{2} \frac{p^{2}_{\rm \vartheta}}{\mu r^2} + \frac{1}{\mu c^2} \frac{1}{1-\frac{2\Phi}{\mathcal{M}c^2} }\times\\ \times \Bigg( \frac{1}{2} \mathcal{E}^2 \frac{\mu^3}{\mu_{\rm 3}^3} 
+\,\mathcal{E}\,\Phi\bigg(\frac{\mu^3}{\mu_{\rm 3}^3} - \frac{\mu}{\mathcal{M}}\bigg)+\frac{1}{2} \Phi^2  \bigg(\frac{\mu^3}{\mu_{\rm 3}^3} - 2\frac{\mu}{\mathcal{M}}\bigg)  \Bigg) \Bigg] \Bigg\}^{-\frac{1}{2}}dr,
\end{gathered}
\end{multline}
where $r_{\rm c}$ is the distance of closest approach, that can be found by setting $dr/d\vartheta =0$ in Eq. (\ref{rc}). It is now useful to make a change of variable, by introducing $x={r_{\rm c}}/{r}$ so that Equation (\ref{net}) becomes
\begin{multline}\label{defl}
\begin{gathered}
\vartheta= 2\int_{0}^{1}{\frac{dx}{x}} \Bigg\{  \bigg[ \frac{2\mu r^{2}_{\rm c}E}{p^{2}_{\rm \vartheta}x^2} + \frac{2\mu^2 G \mathcal{M}r_{\rm c}}{p^{2}_{\rm \vartheta}x^2} \bigg] \bigg(\frac{1}{1-\frac{2G\mu x}{r_{\rm c}c^2}} \bigg)-1\,+ \\
+\frac{2 r^{2}_{\rm c}}{p^{2}_{\rm \vartheta}x^2 c^2} \bigg(\frac{1}{1-\frac{2G\mu x}{r_{\rm c}c^2}} \bigg) 
 \Bigg[\frac{1}{2} \mathcal{E}^2 \frac{\mu^3}{\mu_{\rm 3}^3} + \frac{\mathcal{E}G\mu \mathcal{M}}{r_{\rm c}}x \Bigg(\frac{\mu^3}{\mu_{\rm 3}^3} -\frac{\mu}{\mathcal{M}}  \Bigg) +\\
 +\frac{1}{2} \frac{G^2\mu^2 \mathcal{M}^2 x^2}{r^{2}_{\rm c}} \Bigg(\frac{\mu^3}{\mu_{\rm 3}^3} -2\frac{\mu}{\mathcal{M}}  \Bigg)\Bigg]        
\Bigg\}^{-\frac{1}{2}}.
\end{gathered}
\end{multline}
Furthermore, let ${G\mu}/{r_{\rm c}c^2}={\delta}/{r_{\rm c}}\ll 1$, so that we can perform the expansion 
$(1-{2\delta x}/{r_{\rm c}})^{-1} \approx 1+{2\delta x}/{r_{\rm c}}$ and neglect the terms proportional to $1/c^4$ obtaining
\begin{equation}\label{defl2}
\vartheta= 2\int_{0}^{1}{ \big[-a_{\rm 1}x^2+a_{\rm 2}r_{\rm c}\,x+a_{\rm 3}r^{2}_{\rm c} \big]^{-1/2}} dx,
\end{equation}
where we have defined the following quantities to simplify the notation,
\begin{gather}\label{abccoef}
a_{\rm 1}= 1-\frac{G^2\mu^2 \mathcal{M}^2}{p^{2}_{\rm \vartheta}c^2} \bigg( \frac{\mu^3}{\mu_{\rm 3}^3} + 2\frac{\mu}{\mathcal{M}}\bigg),\nonumber\\
a_{\rm 2}= \frac{2G\mu^2 \mathcal{M}}{p^{2}_{\rm \vartheta}}\bigg\{1+ \frac{\mathcal{E}}{c^2}\bigg[ \frac{1}{\mu}\bigg(\frac{\mu^3}{\mu_{\rm 3}^3} +\frac{\mu}{\mathcal{M}} \bigg)\bigg]  \bigg\},\nonumber\\
a_{\rm 3}= \frac{2\mu E}{p^{2}_{\rm \vartheta}}\bigg(1+\frac{\mathcal{E}}{2\mu c^2}  \frac{\mu^3}{\mu_{\rm 3}^3} \bigg).
\end{gather}
Solving the elementary integral in Eq. (\ref{defl2}), yields
\begin{equation}\label{thetafinale}
\vartheta= \frac{2}{\sqrt{a_{\rm 1}}} \Bigg[\sin^{-1}{\Bigg(\frac{a_{\rm 2}r_{\rm c}}{\sqrt{a_{\rm 2}^2 r^{2}_{\rm c}+4a_{\rm 1}a_{\rm 3}r_{\rm c}^{2}}}\Bigg)}-\sin^{-1}{\Bigg(\frac{a_{\rm 2} r_{\rm c}-2a_{\rm 1}}{\sqrt{a_{\rm 2}^2 r^{2}_{\rm c}+4a_{\rm 1}a_{\rm 3}r_{\rm c}^{2}}} \Bigg)} \Bigg].
\end{equation}
To eliminate the variable $r_{\rm c}$ from the expression above, we solve $\big[-a_{\rm 1}x^2+a_{\rm 2}r_{\rm c}x+a_{\rm 3}r^{2}_{\rm c} \big]_{x=1} = 0$
and substitute its positive root in Eq. (\ref{thetafinale}), getting
\begin{equation}\label{THETA}
\vartheta= \frac{2}{\sqrt{a_{\rm 1}}} \bigg[\sin^{-1}{\bigg(1+\frac{4a_{\rm 1}a_{\rm 3}}{a_{\rm 2}^2}\bigg)^{-1/2}} + \frac{\pi}{2} \bigg].
\end{equation}
The latter is the azimuthal angle of the collision in the frame $\textit{S}^{\prime\prime}$. In this form $\vartheta$ is given as a function of the first integrals $p_{\rm \vartheta}$ and $\mathcal{E}$. We now proceed to express it in terms of the impact parameter $b$ and the asymptotic relative velocity of the stars in the frame $\textit{S}^{\prime\prime}$.\\
\indent Let $\textbf{\textit{u}}$, $\textbf{\textit{u}}_{\rm T}$, $\textbf{\textit{u}}_{\rm F}$ be the asymptotic values of $\textbf{u}$, $\textbf{u}_{\rm T}$, $\textbf{u}_{\rm F}$ in $\textit{S}^{\prime\prime}$ after the encounter. In this limit (i.e. $r\rightarrow+\infty$) we have that $r^2{d\vartheta}/{dt}\rightarrow bu$, $\Phi\rightarrow 0$ and $\mathcal{E}\rightarrow \mu \textit{u}^2/2$, so Equations (\ref{ptheta},\ref{Hdarwin}) become 
\begin{equation}
p_{\rm \vartheta} \approx \mu b \textit{u} \bigg(1+ \frac{\textit{u}^2}{2c^2} \frac{\mu^3}{\mu_{\rm 3}^3}\bigg),\quad E\approx \frac{1}{2} \mu \textit{u}^2  \bigg(1+ \frac{3}{4} \frac{\textit{u}^2}{c^2}\frac{\mu^3}{\mu_{\rm 3}^3}\bigg).
\end{equation}
Introducing the effective impact parameter for sharp deflections $\mathcal{R}= {G\mathcal{M}}/{\textit{u}^2}$, the factors $a_i$ in Eq. (\ref{defl2}) become
\begin{eqnarray} 
a_{\rm 1}= 1-\frac{\textit{u}^2\mathcal{R}^2}{b^2 c^2}\bigg( \frac{\mu^3}{\mu_{\rm 3}^3}+2\frac{\mu}{\mathcal{M}} \bigg),\nonumber\\
a_{\rm 2}= \frac{2\mathcal{R}}{b^{2}}\bigg[1+ \frac{\textit{u}^2}{2 c^2} \bigg(\frac{\mu}{\mathcal{M}}-\frac{\mu^3}{\mu_{\rm 3}^3}\bigg)\bigg],\nonumber\\
a_{\rm 3}= \frac{1}{b^2}.
\end{eqnarray}
The azimuthal angle $\vartheta$ is finally given as a function of $b$ and $\textit{u}$ as 
%\begin{strip}
\begin{multline}\label{azimutale}
\begin{gathered}
\vartheta = \frac{2\Bigg\{ \sin^{-1}{ \Bigg(\frac{  \tiny\frac{4\mathcal{R}^2}{b^4}\bigg[1+\frac{\textit{u}^2}{c^2}\bigg(\frac{\mu}{\mathcal{M}}-\frac{\mu^3}{\mu_{\rm 3}^3}\bigg)\bigg]+\frac{4}{b^2} \bigg[1-\frac{\textit{u}^2}{c^2} \frac{\mathcal{R}^2}{b^2} \bigg(\frac{\mu^3}{\mu_{\rm 3}^3}+2\frac{\mu}{\mathcal{M}}\bigg)\bigg] }{\frac{4\mathcal{R}^2}{b^4} \big\{1+\frac{\textit{u}^2}{c^2}\big[\big(\frac{\mu}{\mathcal{M}}-\frac{\mu}{\mu_{\rm 3}}\big)^3   \big]\big\}}\Bigg)^{-\frac{1}{2}}}
+\frac{\pi}{2} \Bigg\}}{\tiny\sqrt{1-\frac{\textit{u}^2\mathcal{R}^2}{b^2 c^2}\bigg(\frac{\mu^3}{\mu_{\rm 3}^3}+2\frac{\mu}{\mathcal{M}}\bigg) }}\approx\\
\approx \tiny\frac{2 \sin^{-1}{\bigg(1+\frac{b^2}{\mathcal{R}^2}-\frac{\textit{u}^2}{c^2} \bigg[\bigg(\frac{\mu^3}{\mu_{\rm 3}^3}+ 2\frac{\mu}{\mathcal{M}}\bigg)+\frac{b^2}{\mathcal{R}^2}\Big(\frac{\mu}{\mathcal{M}}- \frac{\mu^3}{\mu_{\rm 3}^3}\Big)  \bigg]\bigg)^{-\frac{1}{2}}+\pi }      }{\tiny\sqrt{1-\frac{\textit{u}^2\mathcal{R}^2}{b^2 c^2}\bigg(\frac{\mu^3}{\mu_{\rm 3}^3}+2\frac{\mu}{\mathcal{M}}\bigg) }},
\end{gathered}
\end{multline}
%\end{strip}
from which one may recover the net deflection angle as $\theta_{\rm defl}=\vartheta-\pi$.\\
\indent In order to switch back to $\textit{S}^{\prime}$ and evaluate $\Delta \textbf{v}_{\rm T \parallel}$, let ${\mathbf{u}}^{i}_{\rm T}$ and ${\mathbf{u}}^{f}_{\rm T}$ be the initial and final velocity  of the test star in $\textit{S}^{\prime\prime}$. By decomposing them into the components parallel and perpendicular to ${\textbf{u}}^{i}_{\rm T}$, we have
\begin{gather}
{u}^{i}_{\rm T \parallel} =\frac{ \textbf{\textit{u}}^{i}_{\rm T} \cdot \textbf{\textit{u}}^{i}_{\rm T}}{||\textbf{\textit{u}}_{\rm T} ||} \,\, \\
{u}^{i}_{\rm T \perp} =\frac{|| \textbf{\textit{u}}^{i}_{\rm T} \wedge \textbf{\textit{u}}^{i}_{\rm T}||}{||\textbf{\textit{u}}_{\rm T}||} = 0 \,\,\\
{u}^{f}_{\rm T  \parallel} =\frac{ \textbf{\textit{u}}^{i}_{\rm T} \cdot \textbf{\textit{u}}^{f}_{\rm T}}{||\textbf{\textit{u}}_{\rm T}||}=\textit{u}_{\rm T} \cos(\vartheta-\pi)\,\,\\
{u}^{f}_{\rm T  \perp} =\frac{ ||\textbf{\textit{u}}^{i}_{\rm T} \wedge \textbf{\textit{u}}^{f}_{\rm T}||}{||\textbf{\textit{u}}_{\rm T} ||} =\textit{u}_{\rm T} \sin(\vartheta-\pi)\,\,.
\end{gather}
In the $\textit{S}^{\prime}$ frame, $\Delta \textbf{w}_{\rm T}\equiv \textbf{w}^{f}_{\rm T}$, since $\textbf{w}^{i}_{\rm T}=0$. Moreover, as the test star moves with velocity $-\textbf{u}_{\rm T}$ in $\textit{S}^{\prime\prime}$, we can write in components
\begin{equation}\label{veffe}
\Delta w_{\rm T  \parallel}=\frac{ {u}_{\rm T} \cos(\vartheta-\pi)-\textit{u}_{\rm T} }{1-\frac{{u}^{2}_{\rm T} \cos(\vartheta-\pi)}{c^2}} \approx \big[{u}_{\rm T} \cos(\vartheta-\pi)-{u}_{\rm T}\big] 
\bigg(1+ \frac{{u}^{2}_{\rm T}}{c^2} \bigg),
\end{equation}
and
\begin{multline}\label{vutti}
\begin{gathered}
\Delta w_{\rm T  \perp}=\frac{ {u}_{\rm T} \sin(\vartheta-\pi)\sqrt{1-\frac{{u}^{2}_{\rm T}}{c^2}} }{1-\frac{{u}^{2}_{\rm T} \cos(\vartheta-\pi)}{c^2}}\approx \big[ {u}_{\rm T} \sin(\vartheta-\pi) \big] \bigg( 1-\frac{1}{2} \frac{{u}^2_{\rm T}}{c^2}\bigg)\times\\
\times\bigg(1+ \frac{{u}^{2}_{\rm T}}{c^2} \bigg)  \approx \big[{u}_{\rm T} \sin(\vartheta-\pi) \big] \bigg[1+ \frac{ {u}^2_{\rm T}}{c^2} \cos(\vartheta-\pi)-\frac{1}{2}\frac{{u}^2_{\rm T}}{c^2} \bigg]. 
\end{gathered}
\end{multline}
The two equations above should then be expressed in terms of $\textbf{w}_{\rm F}$, instead of $\mathbf{u}_{\rm T}$ and $\mathbf{u}$. Let us remind that $\mathbf{u}$ is the {\it relative velocity} in $\textit{S}^{\prime\prime}$, as given by Eq.(\ref{vrelrel}), where ${u}_{\rm F}$, ${u}_{\rm T}$ are defined in Eq. (\ref{uf1PN}).\\
\indent Since we only need the relative velocity at 1PN approximation, expanding Eq.(\ref{vrelrel}) an keeping the terms in $1/c^2$ yields
\begin{equation}
u\approx||\textbf{u}_{\rm F}-\textbf{u}_{\rm T}||\bigg(1+\frac{Mm}{\mathcal{M}^2}\frac{u^2}{c^2} \bigg)
\end{equation}
where we replaced $\textit{u}_{\rm F}$, $\textit{u}_{\rm T}$ with their expression given by Eq. (\ref{uf1PN}). At the first order we have $\textit{u}=v_{\rm F}$ and the previous equation becomes
\begin{equation} \label{uvf}
u = \bigg(1+ \frac{mM}{\mathcal{M}^2} \frac{w_{\rm F}^{2}}{c^2} \bigg) w_{\rm F}.
\end{equation}
Making use of the definition of ${u}_{\rm T}$ in the limit $\textbf{r}\rightarrow \infty$ given in  \cite{DD},
\begin{equation}\label{uv}
{u}_{\rm T}= \frac{m}{\mathcal{M}} {u} + \frac{1}{2} \frac{mM(M-m)}{\mathcal{M}^3} \frac{w^{3}_{\rm F}}{c^2}=\frac{m}{\mathcal{M}}\bigg(1+\frac{1}{2} \frac{m}{\mathcal{M}} \frac{w_{\rm F}^{2}}{c^2}  \bigg)w_{\rm F},
\end{equation}
and substituting Equations (\ref{veffe}) and (\ref{uvf}) in Equation (\ref{coefffin}), we obtain the DF formula in $1$PN approximation given in Eq. (\ref{pippo1}).
\section{Post-Newtonian deflection angle}\label{app3}
In order to recover the parallel and perpendicular components of the relative velocity during an encounter, we should first evaluate $\cos(\vartheta-\pi)$ and $\sin(\vartheta-\pi)$ in 1PN approximation. In such limit we have that $\cos ({1}/{c^2}.......) \approx 1$ and $\sin ({1}/{c^2}.......) \approx ({1}/{c^2}.......)$. Using the standard trigonometry and some further algebraic manipulation we define the angle $\vartheta$ by its sine and cosine as
%\begin{strip}
\begin{multline}
\begin{gathered}
\cos(\vartheta-\pi)\approx 1-\frac{2}{1+b^2/\mathcal{R}^2}+\\
-\frac{2\textit{u}^2}{c^2} \frac{\Big(\frac{\mu^3}{\mu_{\rm 3}^3}+2\frac{\mu}{\mathcal{M}}\Big)}{(1+b^2/\mathcal{R}^2)^2}-\frac{2\textit{u}^2 b^2} {c^2\mathcal{R}^2} \frac{\Big(\frac{\mu}{\mathcal{M}}-\frac{\mu^3}{\mu_{\rm 3}^3}\Big)}{(1+b^2/\mathcal{R}^2)^2}+\\
-\frac{2\textit{u}^2 \mathcal{R}}{c^2 b}\frac{\Big(\frac{\mu^3}{\mu_{\rm 3}^3}+2\frac{\mu}{\mathcal{M}}\Big)}{(1+b^2/\mathcal{R})^2}\bigg[\sin^{-1}{\big(1+b^2/\mathcal{R}^2 \big)^{-1/2}}+\pi/2 \bigg],
\end{gathered}
\end{multline}
%\end{strip}
and
\begin{multline}
\begin{gathered}
\sin(\vartheta-\pi)\approx \frac{2b}{\mathcal{R}(1+b^2/\mathcal{R^2})}+\\
-\frac{\textit{u}^2 b}{c^2 \mathcal{R}}\bigg[\frac{\mathcal{R}^2}{b^2} \bigg(\frac{\mu^3}{\mu_{\rm 3}^3}+2\frac{\mu}{\mathcal{M}}\bigg)+\bigg(\frac{\mu}{\mathcal{M}}-\frac{\mu^3}{\mu_{\rm 3}^3}\bigg)\bigg]+\\
+\frac{\textit{u}^2}{c^2}\frac{2b}{\mathcal{R}} \frac{\Big[\Big(\frac{\mu^3}{\mu_{\rm 3}^3}+2\frac{\mu}{\mathcal{M}}\Big)+\frac{b^2}{\mathcal{R}^2} \Big(\frac{\mu}{\mathcal{M}}-\frac{\mu^3}{\mu_{\rm 3}^3}\Big) \Big]}{(1+b^2/\mathcal{R}^2)^2}+\\
+\bigg(1-\frac{2}{1+b^2/\mathcal{R}^2}\bigg)\bigg[\frac{\textit{u}^2 \mathcal{R}^2}{c^2b^2} \bigg(\frac{\mu^3}{\mu_{\rm 3}^3}+2\frac{\mu}{\mathcal{M}}\bigg)\\
\times\bigg(\sin^{-1}{\big(1+b^2/\mathcal{R}^2 \big)^{-1/2}}+\frac{\pi}{2} \bigg)\bigg].
\end{gathered}
\end{multline}
\end{appendix}
\end{document}